\documentclass[twocolumn]{aastex63}

\usepackage{graphicx}
\usepackage{mathtools}
\usepackage{hyperref}

\newcommand\hi{\mbox{\sc Hi}}

    \makeatletter
\def\@fnsymbol#1{\ensuremath{\ifcase#1\or \dagger\or \ddagger\or
   \mathsection\or \mathparagraph\or \|\or **\or \dagger\dagger
   \or \ddagger\ddagger \else\@ctrerr\fi}}
    \makeatother

\received{June 18, 2020}
\accepted{January 20, 2021}
\submitjournal{AJ}

\shorttitle{FLAG \hi~Commissioning}
\shortauthors{Pingel et al.}
\graphicspath{{./}{figures/}}
\begin{document}

\title{Commissioning the $\hi$ Observing Mode of the Beamformer for the Cryogenically Cooled Focal L-band Array for the GBT (FLAG)}

\correspondingauthor{Nickolas Pingel}
\email{Nickolas.Pingel@anu.edu.au}

\author[0000-0001-9504-7386]{N.~M.~Pingel}
\affiliation{Research School of Astronomy and Astrophysics \\
The Australian National University  \\
Canberra, ACT 2611, Australia}
\affiliation{Department of Physics and Astronomy \\
West Virginia University \\
White Hall, Box 6315, Morgantown, WV 26506}
\affiliation{Center for Gravitational Waves and Cosmology \\
West Virginia University \\
Chestnut Ridge Research Building, Morgantown, WV 26505}

\author{D.~J.~Pisano}
\affiliation{Department of Physics and Astronomy \\
West Virginia University \\
White Hall, Box 6315, Morgantown, WV 26506}
\affiliation{Center for Gravitational Waves and Cosmology \\
West Virginia University \\
Chestnut Ridge Research Building, Morgantown, WV 26505}
\affiliation{Adjunct Astronomer at Green Bank Observatory, P.O. Box
2, Green Bank, WV 24944, USA.}

\author{M.~Ruzindana}
\affiliation{Brigham Young University (BYU)
Provo, UT, 84602, USA}

\author{M.~Burnett}
\affiliation{Brigham Young University (BYU)
Provo, UT, 84602, USA}

\author[0000-0002-8043-6909]{K.~M.~Rajwade}
\affiliation{Jodrell Bank Centre for Astrophysics \\
University of Manchester \\
Oxford Road, Manchester M193PL, UK}

\author{R.~Black}
\affiliation{Brigham Young University (BYU)
Provo, UT, 84602, USA}

\author{B.~Jeffs}
\affiliation{Brigham Young University (BYU)
Provo, UT, 84602, USA}
\author{K.~F.~Warnick}
\affiliation{Brigham Young University (BYU)
Provo, UT, 84602, USA}

\author{D.~R.~Lorimer}
\affiliation{Department of Physics and Astronomy \\
West Virginia University \\
White Hall, Box 6315, Morgantown, WV 26506}
\affiliation{Center for Gravitational Waves and Cosmology \\
West Virginia University \\
Chestnut Ridge Research Building, Morgantown, WV 26505}

\author{D.~Anish Roshi}
\affiliation{National Radio Astronomy Observatory (NRAO)\\
520 Edgemont Road Charlottesville, VA 22903, USA} 
\affiliation{Arecibo Observatory \\
Arecibo, Puerto Rico 00612}

\author{R.~Prestage}
\thanks{Deceased}
\affiliation{Green Bank Observatory (GBO) \\
155 Observatory Rd, Green Bank, WV 24944, USA}

\author{M.~A.~McLaughlin}
\affiliation{Department of Physics and Astronomy \\
West Virginia University \\
White Hall, Box 6315, Morgantown, WV 26506}
\affiliation{Center for Gravitational Waves and Cosmology \\
West Virginia University \\
Chestnut Ridge Research Building, Morgantown, WV 26505}

\author{D. Agarwal}
\affiliation{Department of Physics and Astronomy \\
West Virginia University \\
White Hall, Box 6315, Morgantown, WV 26506}
\affiliation{Center for Gravitational Waves and Cosmology \\
West Virginia University \\
Chestnut Ridge Research Building, Morgantown, WV 26505}

\author{T.~Chamberlin}
\affiliation{Green Bank Observatory (GBO) \\
155 Observatory Rd, Green Bank, WV 24944, USA}

\author{L.~Hawkins}
\affiliation{Green Bank Observatory (GBO) \\
155 Observatory Rd, Green Bank, WV 24944, USA}

\author{L.~Jensen}
\affiliation{Green Bank Observatory (GBO) \\
155 Observatory Rd, Green Bank, WV 24944, USA}

\author{P.~Marganian}
\affiliation{Green Bank Observatory (GBO) \\
155 Observatory Rd, Green Bank, WV 24944, USA}

\author{J.~D.~Nelson}
\affiliation{Green Bank Observatory (GBO) \\
155 Observatory Rd, Green Bank, WV 24944, USA}

\author{W.~Shillue}
\affiliation{National Radio Astronomy Observatory (NRAO)\\
520 Edgemont Road Charlottesville, VA 22903, USA}

\author{E. Smith}
\affiliation{Department of Physics and Astronomy \\
West Virginia University \\
White Hall, Box 6315, Morgantown, WV 26506}
\affiliation{Center for Gravitational Waves and Cosmology \\
West Virginia University \\
Chestnut Ridge Research Building, Morgantown, WV 26505}

\author{B.~Simon}
\affiliation{Green Bank Observatory (GBO) \\
155 Observatory Rd, Green Bank, WV 24944, USA}

\author{V. Van Tonder}
\affiliation{Square Kilometre Array South Africa (SKA SA) \\
Cape Town, South Africa}

\author{S.~White}
\affiliation{Green Bank Observatory (GBO) \\
155 Observatory Rd, Green Bank, WV 24944, USA}

\begin{abstract}

We present the results of commissioning observations for a new digital beamforming back end for the Focal plane L-band Array for the Robert C. Byrd Green Bank Telescope (FLAG), a cryogenically cooled Phased Array Feed (PAF) with the lowest measured $T_{\rm sys}$/$\eta$ of any PAF outfitted on a radio telescope to date. We describe the custom software used to apply beamforming weights to the raw element covariances to create research quality spectral line images for the new fine-channel mode, study the stability of the beam weights over time, characterize FLAG's sensitivity over a frequency range of 150 MHz, and compare the measured noise properties and observed distribution of neutral hydrogen emission from several extragalactic and Galactic sources with data obtained with the current single-pixel L-band receiver. These commissioning runs establish FLAG as the preeminent PAF receiver currently available for spectral line observations on the world's major radio telescopes.  

\end{abstract}

%% KEYWORDS

%% INTRODUCTION
\section{Introduction}\label{sec:intro}

The increase in survey speed provided by Phased Array Feed (PAF) receivers embodies the next major advancement in radio astronomy instrumentation. Such arrays have been used commercially for decades \citep{milligan05}, but the unique challenge of operating at extremely low noise levels to detect inherently faint astrophysical signals has only been overcome within the last two decades (e.g.,~\citealt{FisherBradley2000}). Placing an array of densely packed dipole radiators in the focal plane of a radio telescope allows full sampling of the focal field. Multiplying voltages from the dipoles by different complex coefficients (i.e.,~beamformer weights) and summing them will alter the aperture illumination such that the resulting far-field power patterns mimic a multi-beam feed (e.g.~\citealt{Landon10}), while avoiding the challenges of positioning physically distinct feeds. This is an especially powerful shortcut for L-band observations where relatively large physical feeds are necessary and only sample a limiting fraction of sky at one instant. 

Several PAFs have successfully been tested and deployed on both large single dishes, such as the 64m Parkes telescope, and aperture synthesis arrays. For instance, \citet{Reynolds17} successfully recreated a detailed neutral hydrogen ($\hi$) column density (N$_{\rm HI}$) map of the Large Magellanic Cloud, originally observed with the Parkes' L-band multi-beam receiver, as well as the direct detection of source from the $\hi$ Parkes All-Sky Survey (HIPASS; \citealt{barnes01}) and hydrogen recombination lines. \citet{Serra15} utilized the PAF-equipped Australian Square Kilometer Array Pathfinder (ASKAP) to reveal new $\hi$ clouds within the IC 1459 galaxy group. More recently, pilot observations of Widefield ASKAP L-band Legacy All-sky Blind Survey (WALLABY) have expanded the total membership of the NGC 7162 galaxy group and provided high-quality $\hi$ data for kinematic modeling \citep{Reynolds19}, identified five new $\hi$ sources in the NGC 7232 group \citep{kleiner2019}, and characterized $\hi$ clouds that are likely resolved tidal debris features from the NGC 7232/3 triplet \citep{Lee-Waddell19}. Additional early science results from WALLABLY are discussed in \citet{elaglai2019} and \citet{for2019}. Other recent observations from The Galactic ASKAP (GASKAP; \citealt{dickey2013}) survey of the $\hi$ in the nearby Small Magellanic Cloud, where the $\sim$5$\times$5 deg$^2$ extent of the dwarf galaxy was captured in a single pointing, have demonstrated the clear advantage PAFs provide in creating wide-field images \citep{mcclure-griffiths2018}. Additionally, commissioning observations from the Apertif upgrade to the Westerbork Radio Telescope (WSRT; \citealt{oosterloo09}) have shown excellent wide-field imaging capabilities.

While the increase in the Field-of-View (FoV) will in turn dramatically increase the survey speeds of aperture synthesis arrays like the Apertif or ASKAP, the small filling factors and spacing of the individual antenna elements inherently filter out the largest spatial frequencies and limit the sensitivity to low surface brightness emission. Complimentary observations from a large single dish provide these vital missing zero spacing measurements to ensure angular sensitivity at large scales and high surface brightness sensitivity. The decrease in the necessary telescope time required for deep (N$_{\rm HI}$ $\leq$ 10$^{18}$ cm$^{-2}$) on-the-fly (OTF) mapping of extended sources, makes a PAF-equipped GBT the ideal instrument for future deep $\hi$ surveys to reach pioneering sensitivity levels.

%The unblocked aperture design of the 100-m GBT provides a largely unaltered aperture illumination, which facilitates superior sidelobe suppression near the main lobe of the beam. The maximum signal-to-noise ratio (maxSNR) beamformer implementation is designed to optimally illuminate the aperture for each electronically formed beam. However, a maxSNR beamformer gives little control over the sidelobe levels in the resulting far-field beam.  The inherent suppression of sidelobe structure from the unique design of the GBT, combined with the decrease in the necessary telescope time required for deep (N$_{HI}$ $\leq$ 10$^{18}$ cm$^{-2}$) on-the-fly (OTF) mapping of extended sources, makes a PAF equipped GBT the ideal instrument for future deep $\hi$ surveys to reach pioneering sensitivity levels.

The Focal L-band Array for the GBT (FLAG) is a 19 element, dual-polarization PAF with cryogenically cooled low noise amplifiers (LNAs) to maximize sensitivity over a bandwidth of 150.519 MHz divided up into 500 coarse channels. Previous commissioning observations of the front end have shown excellent performance in terms of sensitivity and spectral line imaging capabilities \citep{roshi18}. In Spring 2018, FLAG recorded the lowest reported system temperature ($T_{\rm sys}$) normalized by aperture efficiency $\eta$ at 25$\pm$3 K near 1350 MHz for an electronically formed beam \citep{roshi18}, which is comparable to the capabilities of the existing single-pixel L-band receiver. The work presented in this paper describes aspects of a new digital beamforming back end with a new polyphase filterbank (PFB) implementation for fine channelization of 100 coarse channels into 3200 fine-channels specifically designed for spectral line science. \citet{rajwade2019} provides an overview of the real-time beamforming mode for the detection of transient signals from fast radio bursts and pulsars. 

We describe the system architecture and available observing modes in Section~\ref{sec:FLAGBackground} and briefly summarize the mathematical principles of beamforming in Section~\ref{sec:snrtheory}; in Section~\ref{sec:obs}, we describe the observing setup and strategies for beamformer weight calibration and $\hi$ mapping with the GBT; the custom software used for post-correlation beamforming, flux calibration, and imaging are summarized in Section~\ref{sec:dataReduction}; Section~\ref{sec:results} investigates how distinct sets of beamforming weights vary with time, demonstrates the sensitivity across the full range of bandwidth, compares the $\hi$ properties of several extragalactic and Galactic sources as detected by FLAG and the current L-band single-pixel receiver, and presents a comparison between the survey speed of FLAG relative to other PAFs and multi-beam receivers equipped on the world's major radio telescopes; finally, our conclusions and instrument outlook are summarized in Section~\ref{sec:conc}.

\section{FLAG System Architecture}\label{sec:FLAGBackground}

The Focal L-band Array for the Green Bank Telescope (FLAG) was developed in collaboration between the National Radio Astronomy Observatory (NRAO), the Green Bank Observatory (GBO), Brigham Young University (BYU), and West Virginia University (WVU). It is a 19 element, dual-polarization, cryogenic PAF with direct digitization of radio frequency (RF) signals at the front end, digital signal transport over fiber, and now possesses a real-time signal processing back end with up to 150 MHz bandwidth. The front end employs a new digital-down-link (DDL) mechanism that performs all analog-to-digital conversions in a compact assembly that sits at prime focus \citep{morgan2013unformatted}. 

Two integral processes in the success of the DDL are achieving bit and byte lock in the back end system. The front end system produces complex sample voltages for each dipole element that are serialized into 8-bit real and 8-bit imaginary components. These are combined to form a 16-bit (or 2-byte) word per time sample. These serialized voltages are transmitted over optical fiber without any sort of encoding such as start/stop bits to delineate the boundaries between bits. Bit lock refers to the recovery of the most-significant bit by the deserializer in the FLAG back end. This is done by constructing a histogram of the arriving samples and comparing to the expected probability density function of a random Gaussian process. Once the sample are correctly aligned in terms of their most-significant bits, the byte-lock procedure ensures that two sequential bytes are correctly identified as the real and imaginary components. Due to the relationship between the magnitudes of complex conjugated signals, if the bytes are incorrectly identified (i.e., there is no byte-lock), a strong test-tone injected at a known positive frequency offset relative to a set central frequency will have a symmetric counterpart at the corresponding negative frequency offset. The bits are then slipped by eight locations to correctly align the bytes to achieve byte-lock. See \citet{junming2017} and \citet{burnett2017masters} for detailed information on the PAF receiver front end and bit/byte locking procedures, respectively.

The FLAG back end consists of five digital optical receiver cards, five ROACH II FPGA boards \citep{parsons2006petaop}, a Mellanox SX 1012 12-port 40 Gbe ethernet switch, and five Mercury GPU408 4U GPU Server High Performance Computers (HPCs). These parts are all connected in the order listed.

The digitized signals from the front end of the system are serialized and sent over 40 (38 + 2 spare) optical fibers to the optical receiver cards which are connected to the ROACH II boards. The boards channelize the approximately 150 MHz bandwidth into 512 channels each with a bandwidth of 303.18 kHz. 

The data is then reduced to 500 frequency channels and packetized into 10 user-datagram protocol (UDP) packets each containing 50 frequency samples for eight antennas across 20 time samples. These packets are streamed over 10-Gbe/40-Gbe breakout cables into a 12-port 40-Gbe network switch, which redirects packets into the HPCs such that each one receives 100 frequency samples with a width of 303.18 kHz for all 40 antennas.

\begin{table*}
\resizebox{\textwidth}{!}
{\begin{tabular}{lllll}
\hline \hline
Mode    & Bandwidth {[}MHz{]}  & N$_{\rm chan}$ & N$_{\rm chan}$ in Bank & $\Delta\nu$ {[}kHz{]} \\
\hline
CALCORR & 151.59               & 500      & 25 non-contiguous      & 303.18                \\
PFBCORR & 30.318               & 3200     & 160 contiguous      & 9.47                  \\
RTBF    & 151.59               & 500      &  25 non-contiguous  & 303.18 \\               
\hline
\end{tabular}}\caption {Properties of Available FLAG Observing Modes}
\label{tab:modeSummary} 
\end{table*}

Each HPC then takes these 100 frequency samples and divides them evenly between two Nvidia GeForce Titan X Graphical Processing Units (GPUs), which contain real-time beamformer and coarse/fine channel correlator algorithms. Within each HPC is a real-time operating system (RTOS) called HASHPIPE used for thread management and pipelining, and a user interface called dealer/player. These enable the operation of the beamformer and correlator algorithms. Each HPC can be run in three distinct observing modes: (1) CALCORR, which is the mode used to derive the beamforming weights; (2) PFBCORR, which is used for the spectral line observations and sends a frequency chunk of 100 coarse channels with a total bandwidth of 30.318 kHz through a polyphase filterbank implementation to obtain 3200 total fine channels with resolution of 9.47 kHz; each GPU in these correlator modes runs a correlator thread that processes one-tenth the total bandwidth; and (3) RTBF mode, which is the mode used for pulsar and transient detection. The properties of these observing modes are summarized in Table~\ref{tab:modeSummary}. We refer the reader to \citet{mark2017beamform} for a detailed description on the FLAG back end and \citet{rajwade2019} for the description and early success of the RTBF mode.

%% SECTION OUTLINING BEAMFORMING THEORY 
\section{Maximum Signal-to-Noise Beamforming}\label{sec:snrtheory}
The process of beamforming involves the weighted sum of the individual sensor responses to an incident astronomical signal. In radio astronomy, where the signals are inherently extremely faint, it is advantageous for an observer to compute weights that maximize the signal-to-noise from a given detection. Defining $\mathbf{z}\left(t\right)$ to be a vector containing the individual responses of each dipole in an $M$-dipole PAF measured over a discrete time sample (i.e., integration), a convenient covariance matrix
%% EQUATION FOR COVARIANCE
\begin{equation}\label{eq:covariance}
\mathbf{R} = \mathbf{z}^H\left(t\right)\mathbf{z}\left(t\right)
\end{equation} 
can be constructed such that \textbf{R} is a $M\times M$ matrix of complex values that  characterizes the correlations between the recorded complex voltages of the individual dipole elements. Note that the $H$ superscript in the above equation represents the Hermitian (complex conjugate transpose) form of the vector. 
\citet{jeffs2008} goes on to characterize the signal from the array by the equation
%% EQUATION FOR SIGNAL + NOISE COVARIANCE
\begin{equation}\label{eq:R}
\mathbf{R} = \mathbf{R_{\rm s}} + \mathbf{R_{\rm n}},
\end{equation}
where \textbf{R}$_{\rm s}$ is the signal covariance matrix and \textbf{R$_{\rm n}$} contains the noise covariance from spillover, background, and the mutual coupling of the dipoles.

\textbf{R$_{\rm n}$} can be measured by pointing the telescope to a blank patch of sky so that \textbf{R} $\approx$ \textbf{R}$_{\rm n}$. Pointing at a bright point source and solving Equation~\ref{eq:R} for \textbf{R}$_{\rm s}$ gives the signal covariance matrix. A steering vector that characterizes the response of each dipole in a given direction can now be computed and is defined by
%% EQUATION FOR EXPLICIT DERIVATION OF THE ARRAY RESPONSE
\begin{equation}\label{eq:steerVec}
\mathbf{a}\left(\theta\right) = \mathbf{R_{\rm n}}\mathbf{u}_{\textrm{max}}, 
\end{equation}
where \textbf{u}$_{\rm max}$ is the dominant eigenvector of the generalized eigenvalue equation \textbf{R}\textbf{u}$_{\rm max}$ = $\lambda_{\rm max}$\textbf{R}$_{\rm n}$\textbf{u}$_{\rm max}$. 

\citet{Elmer12} define the maximum signal-to-noise beamformer by maximizing the following expression
%EQUATION FOR SMR
\begin{equation}\label{eq:snr}
\mathbf{w_{\rm maxSNR}} = \mathrm{argmax}\left( \frac{\mathbf{w^H}\mathbf{R_{\rm s}\mathbf{w}}}{\mathbf{w^H}\mathbf{R_{\rm n}\mathbf{w}}}\right).
\end{equation}
The values contained within the weight vector \textbf{w} and its Hermitian form are not yet known. Maximizing Equation~\ref{eq:snr} by taking the derivative with respect to \textbf{w} and setting the result equal to zero is equivalent to finding the dominant eigenvector of the generalized eigenvalue equation
%% EQUATION FOR MAX EIGENVALUE
\begin{equation}\label{eq:eigen}
\mathbf{R_{\rm s}}\mathbf{w_{\rm maxSNR}} = \lambda_{\rm max}\mathbf{R_{\rm n}}\mathbf{w_{\rm maxSNR}}.
\end{equation}
A raw power value $P$ in units of counts at a particular frequency $\nu$ and short term integration ($n$) is measured by calculating
%% EQUATION FOR BEAMFORMED POWER LEVELS
\begin{equation}\label{eq:beamformedPower}
P_{\nu\rm,n} = \mathbf{w^{\rm H}_{\rm maxSNR, \nu\rm,n}}\mathbf{R_{\rm s, \nu\rm, n}}\mathbf{w_{\rm maxSNR, \nu\rm, n}}.
\end{equation}
The max-SNR beamforming algorithm effectively manipulates the individual dipole illumination patterns such that the aperture is optimally illuminated for each formed beam in a given direction on the sky. While this scheme produces the highest gain in a given direction, there is little control over the level of the sidelobes due to the sharp transition in illumination pattern. High sidelobe levels could introduce stray radiation, where signal is detected in a sidelobe rather than the main formed beam, affecting the accuracy of flux and mapped structure. 
For example, stray radiation in the initial data release of the Parkes Galactic All-Sky Survey (GASS; \citet{mcclure-griffiths2009}) accounted for upwards of 35\% of the observed emission in some individual spectra. Nevertheless, high sensitivity over a large field of view is particularly advantageous for the detection of diffuse (angularly extended and faint) $\hi$, as evidenced by the abundance of highly detailed and faint structure observed in the GASS survey even before the application of corrections for stray radiation. The unique unblocked aperture design of the GBT ensures inherently low sidelobe structure --- even in the case of maxSNR --- and subsequently high image fidelity. A PAF-equipped GBT will produce high quality maps while also decreasing the survey times necessary to pursue --- amongst many applications ---  the detection of cold gas accretion, the study of high velocity clouds (\citealt{moss13}), and the compact clouds being driven from the Galactic center (\citealt{diTeodoro18}).

%% SECTION ON OBSERVATIONS
\section{Observations}\label{sec:obs}
The first step in forming beams is the characterization of the response of each individual dipole element in a given direction, $\theta_{\rm i}$, in the form of a signal response vector (i.e., Equation~\ref{eq:steerVec}). For these commissioning observations, we implement a maxSNR beamformer as defined in Equation~\ref{eq:snr}. While a PAF can theoretically form any number of beams as long as there exists a sufficient number of steering vectors and recorded covariance matrices, we employ two calibration techniques deemed a Calibration Grid and 7-Point Calibration to form seven total beams arranged such that the central (i.e., boresight) beam is surrounded by six outer beams in a hexagonal pattern that overlap at approximately the half-power points (see Figure 1 of \citealt{rajwade2019}). This particular pattern provides ideal balance between mapping speed and uniform sensitivity within FLAG's FoV. We refer to the boresight beam as `Beam 0'; as viewed on the sky, Beam 1 is the upper left beam, and the subsequent beam numbers increase in a clockwise fashion. Once a set of \textbf{w}$_{\rm b}$, is obtained for the $b$th beam (of $B$ total beams) in the direction of $\theta_{\rm i}$, we acquire the raw power value at each $\nu$ and short term integration $n$ through Equation~\ref{eq:beamformedPower}. Table~\ref{tab:obsSummary} summarizes all calibration and science observations discussed in this paper.

%% TABLE SUMMARIZING OBSERVATIONS
\begin{table*}
\centering
\resizebox{\textwidth}{!}
{\begin{tabular}{lccccccccc}
    \hline \hline
     \\[-1.0em]
    Session & UT Date & UT Start & UT End & Schedule Block Type & Source & Mode & Integration Length [s] & Central Frequency [MHz] & Notes \\
\hline \\[-0.75em]
%% AGBT16B_400_03
\textbf{GBT16B\_400\_03} & & & & & & & & &\\
 & 2017-05-27 & 04:17:55 & 05:12:11 & Calibration Grid  & 3C295 & CALCORR & 0.1 & 1450.00000 & Continuous Trajectory \\
%% AGBT16B_400_09
\textbf{GBT16B\_400\_09} & & & & & & & & &\\
 & 2017-07-28 & 05:06:19 & 05:38:52 & Calibration Grid  & 3C295 & CALCORR & 0.5 & 1450.00000 & --- \\
%% AGBT16B_400_12
\textbf{GBT16B\_400\_12} & & & & & & & & &\\
 & 2017-08-04 & 04:16:54 & 05:03:33 & Calibration Grid  & 3C295 & CALCORR & 0.5 & 1450.00000 & 40$\times$40 $\square'$ \\
 & {\ddag}2017-08-04  & 05:30:27 & 06:02:27 & DecLatMap  & NGC6946 & PFBCORR & 0.5 & 1450.00000 & 41 columns; $N_{\rm ints}=72$; $t_{\rm eff, comb}=60$ s \\
 & 2017-08-04 & 06:12:19 & 06:14:53 & 7Pt-Calibration & 3C48 & CALCORR & 0.5 & 1450.0000 & 10 s Tracks \\
%% AGBT16B_400_13
\textbf{GBT16B\_400\_13} & & & & & & & & &\\
 & 2017-08-04 & 13:44:40 & 14:29:09 & Calibration Grid  & 3C123 & CALCORR & 0.5 & 1449.84841 & --- \\
% & 2017-08-04 & 14:42:46 & 15:14:47 & DecLatMap & NGC6946 & PFBCORR & 0.5 & 1449.84841 & 41 columns \\
 & 2017-08-04 & 06:12:19 & 06:14:53 & 7Pt-Calibration & 3C134 & CALCORR & 0.5 & 1449.84841 & 15 s Tracks \\
%% AGBT16B_400_14
\textbf{GBT16B\_400\_14} & & & & & & & & &\\
& 2017-08-06 & 16:41:15 & 16:43:58 & 7Pt-Calibration & 3C147 & CALCORR & 0.5 & 1450.0000 & 15 s Tracks \\
& 2017-08-06 & 16:44:48 & 17:22:16 & Calibration Grid  & 3C147 & CALCORR & 0.5 & 1449.74271 & --- \\
%% AGBT17B_360_01
\textbf{GBT17B\_360\_01} & & & & & & & & & \\
 & 2018-01-27 & 15:07:59 & 15:09:55 & 7Pt-Calibration & 3C295 & CALCORR & 0.5 & 1450.0000 & 10 s Tracks \\
  & 2018-01-27 & 15:11:00 & 15:39:18 & Calibration Grid & 3C295 & CALCORR & 0.5 & 1450.0000 & --- \\
 & 2018-01-27 & 15:40:29 & 15:42:24 & 7Pt-Calibration & 3C295 & CALCORR & 0.5 & 1450.00000 & 10 s Tracks \\
%% AGBT17B_360_02
\textbf{GBT17B\_360\_02} & & & & & & & & & \\
 & 2018-01-27 & 18:32:59 & 18:36:07 & 7Pt-Calibration & 3C295 & CALCORR & 0.5 & 1450.00000 & 10 s Tracks;  \\
 & 2018-01-27 & 19:13:57 & 19:41:40 & Calibration Grid & 3C147 & CALCORR & 0.5 & 1450.00000 & --- \\
 & 2018-01-27 & 21:07:00 & 21:10:04 & 7Pt-Calibration & 3C147 & CALCORR & 0.5 & 1450.00000 & 10 s Tracks \\
 %% AGBT17B_360_03
\textbf{GBT17B\_360\_03} & & & & & & & & &\\
 & 2018-01-28 & 06:44:29 & 06:47:38 & 7Pt-Calibration & 3C295 & CALCORR & 0.5 & 1449.84841 & 10 s Tracks \\
 & 2018-01-28 & 06:48:56 & 07:17:23 & Calibration Grid & 3C295 & CALCORR & 0.5 & 1449.84841 & --- \\
 & {\ddag}2018-01-28  & 08:05:49 & 08:36:44 & DecLatMap & NGC4258 Field & PFBCORR & 0.5 & 1449.84841 & 31 columns;$N_{\rm ints}=72$; $t_{\rm eff, comb}=68$ s \\
 & {\ddag}2018-01-28  & 08:05:49 & 08:36:44 & DecLatMap & NGC4258 Field & PFBCORR & 0.5 & 1449.84841 & 31 columns;$N_{\rm ints}=72$; $t_{\rm eff, comb}=68$ s \\
 & {\ddag}2018-01-28  & 08:38:28 & 09:07:35 & DecLatMap & NGC4258 Field & PFBCORR & 0.5 & 1449.84841  & 31 columns;$N_{\rm ints}=72$; $t_{\rm eff, comb}=68$ s \\
%% AGBT17B_360_04
\textbf{GBT17B\_360\_04} & & & & & & & & &\\
 & 2018-01-29 & 07:29:58 & 08:32:14 & Calibration Grid & 3C295 & CALCORR & 0.5 & 1450.00000 & --- \\
& 2018-01-29 & 08:38:51 & 08:42:10 & 7Pt-Calibration & 3C295 & CALCORR & 0.5 & 1450.00000 & 20 s Tracks \\
 & {\ddag}2018-01-29 & 08:50:26 & 09:20:42 & DecLatMap& NGC4258 Field & PFBCORR & 0.5 & 1450.0000 & 31 columns;$N_{\rm ints}=72$; $t_{\rm eff, comb}=68$ s;  \\
 & {\ddag}2018-01-29 & 09:25:19 & 09:56:10 & DecLatMap & NGC4258 Field & PFBCORR & 0.5 & 1450.0000 & 31 columns;$N_{\rm ints}=72$; $t_{\rm eff, comb}=68$ s \\
 & {\ddag}2018-01-29 & 09:59:00 & 10:28:50 & DecLatMap & NGC4258 Field & PFBCORR & 0.5 & 1450.00000 & 31 columns;$N_{\rm ints}=72$; $t_{\rm eff, comb}=68$ s \\
 & {\ddag}2018-01-29 & 10:30:44 & 10:59:11 & DecLatMap & NGC4258 Field & PFBCORR & 0.5 & 1450.00000 & 31 columns;$N_{\rm ints}=72$; $t_{\rm eff, comb}=68$ s \\
%% AGBT17B_360_05
\textbf{GBT17B\_360\_05} & & & & & & & & \\
& 2018-01-30 & 12:02:53 & 12:13:08 & 7Pt-Calibration & 3C295 & CALCORR & 0.5 & 1450.0000 & 20 s Tracks \\
& 2018-01-30 & 12:53:24 & 13:00:44 & 7Pt-Calibration & 3C295 & CALCORR & 0.5 & 1450.00000 & 20 s Tracks \\
%% AGBT17B_360_06
\textbf{GBT17B\_360\_06} & & & & & & & & &\\
& 2018-02-03 & 17:30:03 & 17:35:46 & 7Pt-Calibration & 3C48 & CALCORR & 0.5 & 1075.00000 & 30 s Tracks \\
& 2018-02-03 & 18:15:50 & 18:21:39 & 7Pt-Calibration & 3C48 & CALCORR & 0.5 & 1250.00000 & 30 s Tracks\\
& 2018-02-03 & 18:32:32 & 18:38:21 & 7Pt-Calibration & 3C48 & CALCORR & 0.5 & 1350.00000 & 30 s Tracks \\
& 2018-02-03 & 18:51:01 & 18:56:52 & 7Pt-Calibration & 3C48 & CALCORR & 0.5 & 1550.00000 & 30 s Tracks \\
& 2018-02-03 & 19:08:18 & 19:14:11 & 7Pt-Calibration & 3C48 & CALCORR & 0.5 & 1650.00000 & 30 s Tracks\\
& 2018-02-03 & 19:25:22 & 19:31:17 & 7Pt-Calibration & 3C48 & CALCORR & 0.5 & 1750.00000 &30 s Tracks\\
& 2018-02-03 & 19:57:26 & 20:03:28 & 7Pt-Calibration & 3C48 & CALCORR & 0.5 & 1449.74271 & 30 s Tracks \\
 & 2018-02-03 & 20:04:47 & 20:35:45 & Calibration Grid & 3C48 & CALCORR & 0.5 & 1449.74271 & --- \\
%% AGBT17B_360_07
\textbf{GBT17B\_360\_07} & & & & & & & & &\\
 & 2018-02-05 & 06:25:20 & 06:53:49 & Calibration Grid & 3C295 & CALCORR & 0.5 & 1450.00000 & --- \\
& 2018-02-05 & 10:04:05 & 10:14:38 & 7Pt-Calibration & 3C295 & CALCORR & 0.5 & 1450.00000 & 60 s Tracks \\
%% AGBT17B_455_01
\textbf{GBT17B\_455\_01} & & & & & & & & &\\
& 2018-02-04 & 13:18:07 & 13:28:01 & 7Pt-Calibration & 3C348 & CALCORR & 0.5 & 1450.00000 & 60 s Tracks \\
& {\ddag}2018-02-04 & 13:35:47 & 14:54:23 & RaLongMap & Galactic Center & PFBCORR & 0.5 & 1450.00000 & 41 rows;$N_{\rm ints}=72$; $t_{\rm eff, comb}=60$ s \\
& 2018-02-04 & 15:09:06 & 15:18:56 & 7Pt-Calibration & 3C348 & CALCORR & 0.5 & 1450.84841 & 60 s Tracks \\
& {\ddag}2018-02-04 & 15:26:57 & 16:29:32 & RaLongMap & Galactic Center & PFBCORR & 0.5 & 1450.84841 & 41 rows;$N_{\rm ints}=72$; $t_{\rm eff, comb}=60$ s \\
\hline
\end{tabular}}
\caption {Summary of FLAG Observations; {\ddag} represents mapping scans used to make the science maps; $N_{\rm ints}$ represents the number of integration along each row/column; and $t_{\rm eff,map}$ gives the effective integration time of the combined map time in units of s (see text in Section~\ref{subsec:obsHI}).}
\label{tab:obsSummary} 
\end{table*}

%% SUBSECTION ON CALIBRATION OBSERVATIONS
\subsection{Calibration Grid}\label{subsec:calGrid}

%% FIGURE SHOWING EXAMPLE CALIBRATION GRID
\begin{figure}
\includegraphics[width=3.25 in]{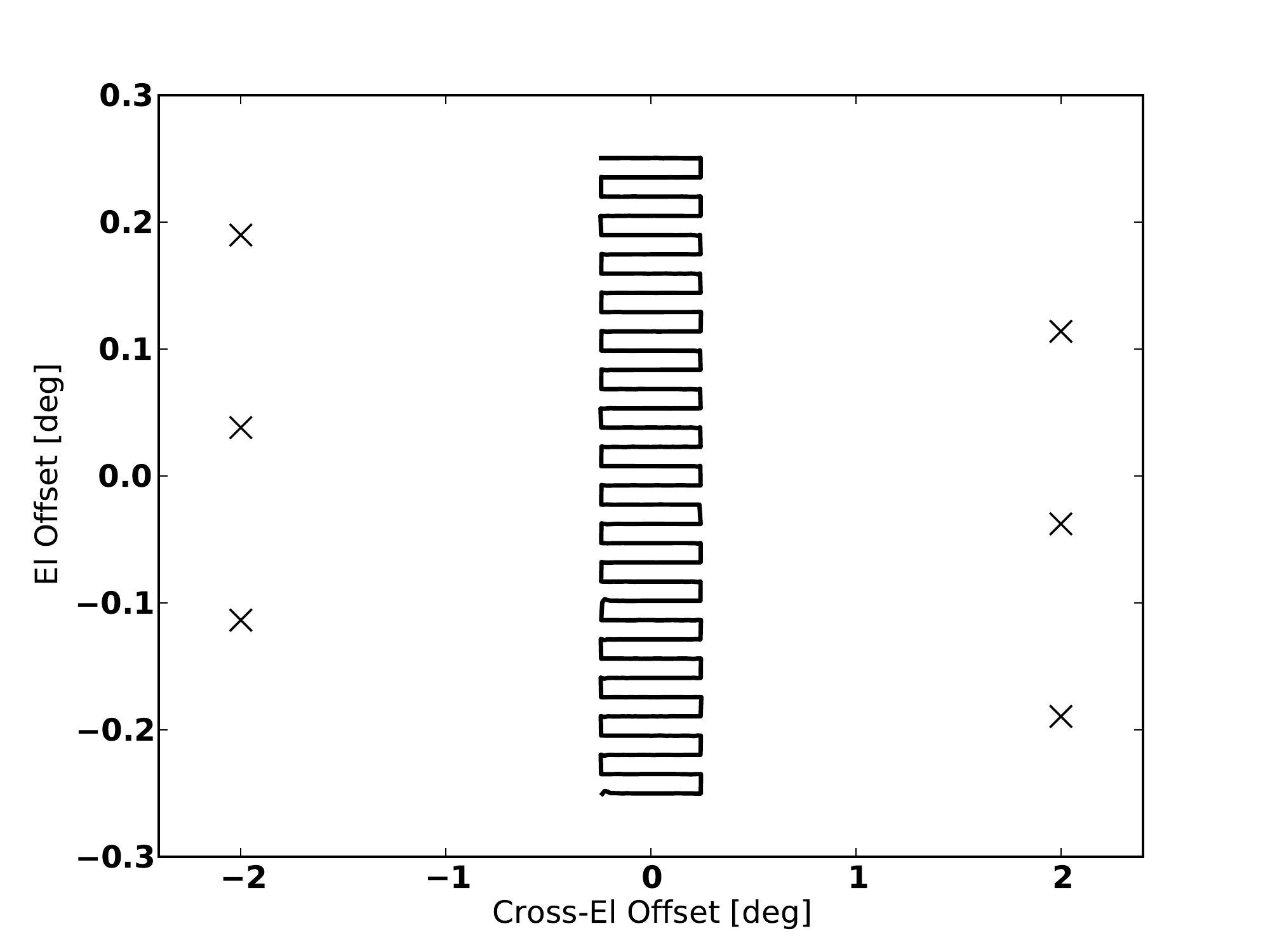}
\caption{\label{fig:calibGridTrajectory} The trajectory from one of our calibration grids centered on 3C295. The `$\times$' symbols denote the mean location of the reference pointings, and the solid black lines represent the trajectory of the grid.}
\end{figure}

To obtain measurements of \textbf{R}$_{\rm s}$, we move the GBT in a grid centered on a strong calibrator spanning 30 arcminutes in Cross-elevation (XEL) as set by the horizontal celestial coordinate system (i.e. `Encoder' setting when using the GBT) for a total of 34 rows spaced 0.91 arcminutes (approximately one-tenth the full-width half max of the GBT beam at 1.4 GHz) apart in Elevation (EL). We compute \textbf{R}$_{\rm n}$ by tracking two degrees in XEL away from the grid for a duration of ten seconds. We track after every fifth row to attain six total reference pointings with three evenly spaced on each side of the grid. To ensure adequate spatial sampling, we move the telescope at a rate of 0.91 arcminutes per second and dump integrations to disk every 0.5 s. The trajectory of the calibration grid observations performed during session GBT16B\_400\_12 centered on 3C295 is shown in Figure~\ref{fig:calibGridTrajectory}. The total time to complete such a grid is about 40 minutes, including scan overhead. 

The calibration grid provides the necessary covariance matrices with which to characterize the response and quality of the formed beams. A convenient quantity with which to compare beam-to-beam sensitivity variations --- as it directly measurable --- is the system equivalent flux density (SEFD), which is the flux density equivalent of the system temperature, $T_{\rm sys}$. The SEFD is defined
%% EQUATION FOR SEFD
\begin{equation}\label{eq:SEFD}
{\rm SEFD} = \frac{S_{\rm CalSrc}}{\left(\frac{\left<P_{\rm s}\right>}{\left<P_{\rm n}\right>} - 1\right)},
\end{equation}
where $S_{\rm CalSrc}$ is the known flux density of a calibrator source in units of Jy and $\left<P_{\rm s}\right>$ and $\left<P_{\rm n}\right>$ are respectively the mean on-source and off-source power values. These are determined by building distributions of on-source and off-source raw beamformed power values contained between coarse channels corresponding to 1400.2 MHz to 1416.6 MHz and 1425.1 MHz to 1440.3 MHz to avoid bias from Galactic $\hi$ emission. These distributions are then fit with separate Gaussian functions to calculate $\left<P_{\rm s}\right>$ and $\left<P_{\rm n}\right>$. The associated uncertainties are taken to be the standard deviations returned by these fits. In cases where the fit does not converge due to complex bandpass shapes, the arithmetic mean and standard deviations are used. All power values are corrected for atmospheric attenuation. The final uncertainty for the SEFD value is computed by propagating the statistical uncertainties of $\left<P_{\rm s}\right>$, $\left<P_{\rm n}\right>$, and $S_{\rm CalSrc}$. The flux density of a given calibrator source is taken from \citet{perleyButler17}.

The SEFD provides a comparison metric between individual beams. If the SEFD is derived for the ideal observation of a blank sky, it can be related to the ratio of $T_{\rm sys}$ and aperture efficiency $\eta$ through 
%% EQUATION FOR Ta/eta
\begin{equation}\label{eq:Ta_eta}
\frac{T_{\rm sys}}{\eta} = \frac{10^{-26}\rm SEFD A_{\rm g}}{2k}, 
\end{equation}
where $A_{\rm g}$ is the geometric area of the GBT, and $k$ is the Boltzmann constant. Substituting the definition for the SEFD from Equation~\ref{eq:SEFD} and putting the power levels in terms of the product between correlation matrices and beamforming weights from Equation~\ref{eq:beamformedPower} results in the expression 
%% FINAL EQUATION FOR TSYS/ETA
\begin{equation}\label{eq:Tsys_eta}
\frac{T_{\rm sys}}{\eta} = \frac{10^{-26}S_{\rm CalSrc}A_{\rm g}}{2k}\frac{\mathbf{w^{\rm H}}\mathbf{R_{\rm n}}\mathbf{w}}{\mathbf{w^{\rm H}}\mathbf{R_{\rm s}}\mathbf{w}}. 
\end{equation}
This equation is an oft-used metric for comparing and characterizing the performance of PAFs \citep{jeffs2008,Landon10, roshi18}, since it can be directly measured. Equation~\ref{eq:Ta_eta} can be rearranged to define a formed beam sensitivity in units of m$^2$ K$^{-1}$ at each $\nu$ from each direction $\theta$
%% EQUATION FOR BEAM SENSITIVITY
\begin{equation}\label{eq:sens}
S_{\nu}\left(\theta\right) = \frac{\eta A_{g}}{T_{\rm sys}} = \frac{2k}{10^{-26}S_{\rm CalSrc}}\frac{\mathbf{w^{\rm H}}\mathbf{R_{\rm s}}\mathbf{w}}{\mathbf{w^{\rm H}}\mathbf{R_{\rm n}}\mathbf{w}}.
\end{equation}

%% FIGURE SHOWING SENSITIVITY MAP
\begin{figure}
\includegraphics[width=\columnwidth]{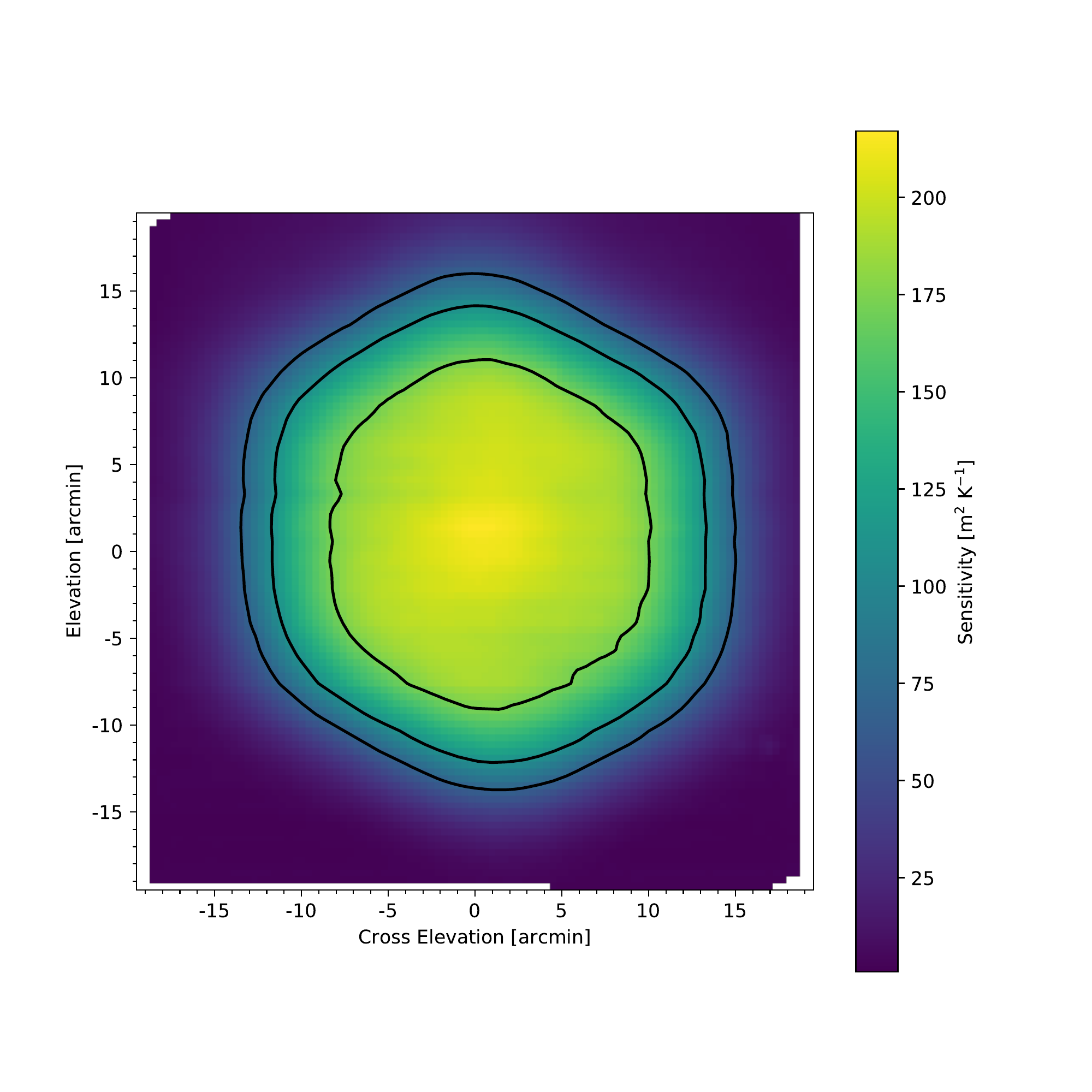}
\caption{Sensitivity map of the XX polarization at 1404.74 MHz derived from the calibration grid shown in Figure~\ref{fig:calibGridTrajectory}. The contours levels begin at the $-$5 dB drop off level of the peak response and continue to the $-$3 and $-$1 dB drop off.}
\label{fig:sensMap}
\end{figure}

%% FIGURE SHOWING BEAM MAP & PROFILES
\begin{figure*}
\includegraphics[width=3.5 in]{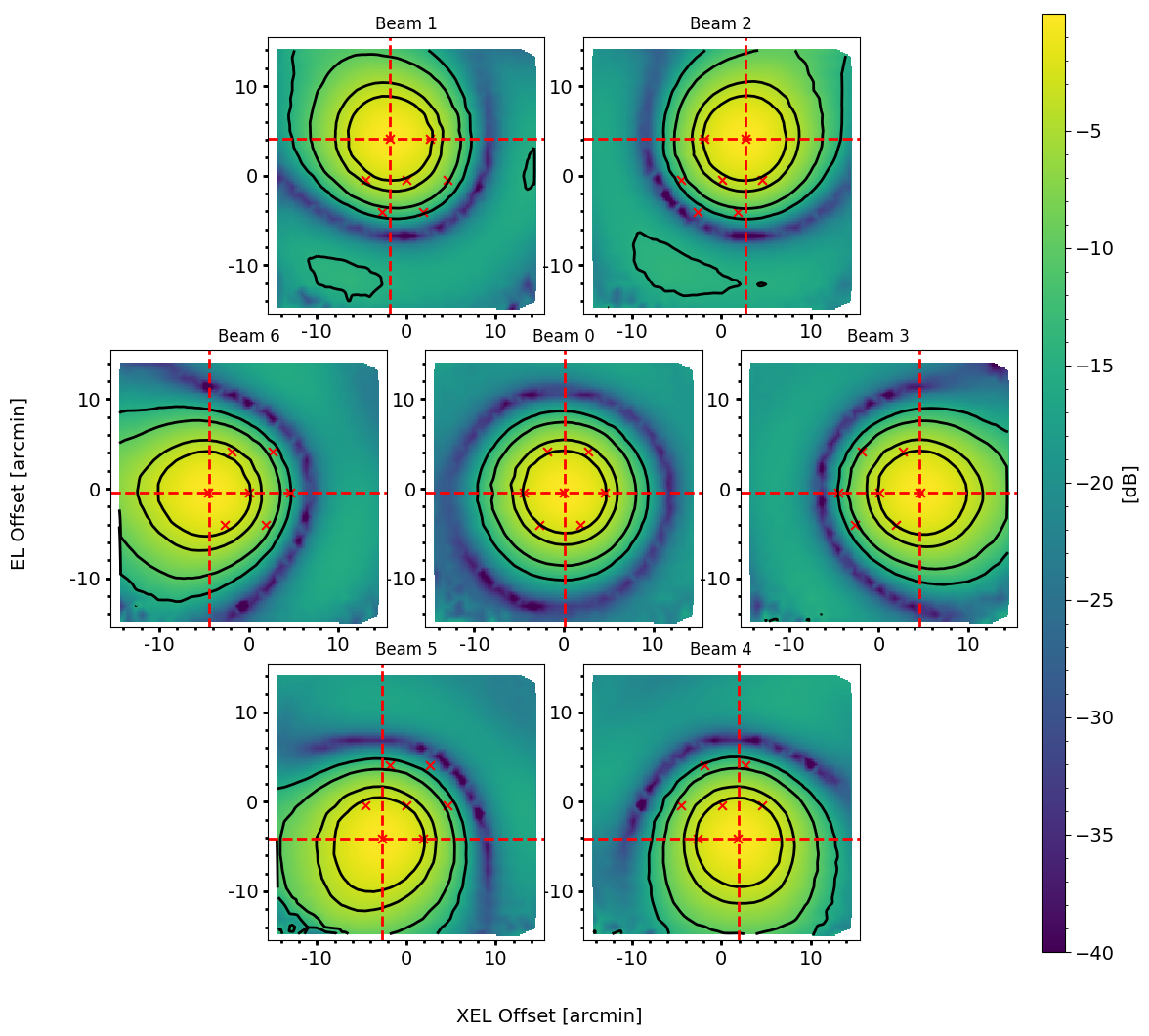}
\includegraphics[width=3.5 in]{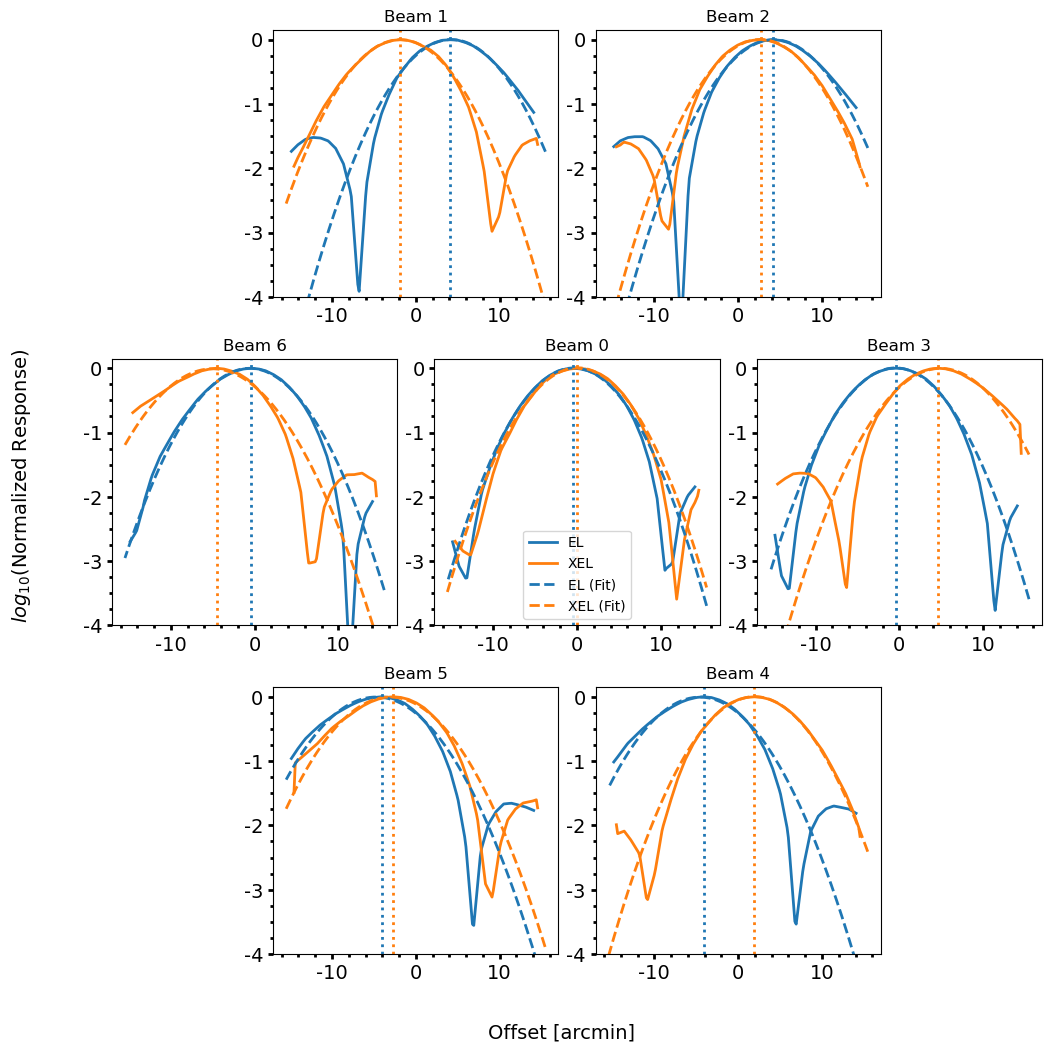}
\caption{left: The formed beam pattern derived from the calibration grid shown in Figure~\ref{fig:calibGridTrajectory}. The red x symbols denote the intended beam centers. The intersections of the vertical and horizontal dashed red lines denote the location of the peak response of each formed beam. The contours represent levels of $-$3, $-$5, $-$10, and $-$15 dB. Right: profiles of the normalized beam response at the location of the peak response along the XEL (orange) and EL (blue) axes. Gaussian fits are represented by dashed lines, while the intended locations of the peak response in XEL and EL are shown by vertical dotted lines.}
\label{fig:beamMap}
\end{figure*}

Figure~\ref{fig:sensMap} shows the resulting sensitivity map from the calibration grid in Figure~\ref{fig:calibGridTrajectory} in the XX pol at 1404.74 MHz. The inner 0.5$\times$0.5 deg$^2$ of the FoV shows uniform sensitivity, which reflects the aggregate response of the individual dipoles on the sky (see Figures 4, 5, and 10 from \citealt{roshi18}), before smoothly dropping towards the edge of the FoV. The excellent uniformity across the FoV facilitates high-quality beams. 

\begin{table*}
\resizebox{\textwidth}{!}
{\begin{tabular}{lcccccccc}
\hline \hline
     \\[-1.0em]
Beam & FWHM$_{\rm XEL}$ [$'$]     & Peak$_{\rm XEL,Intended}$ [$'$] & Peak$_{\rm XEL,Measured}$ [$'$] & XEL \%-Diff & FWHM$_{\rm EL}$ [$'$]      & Peak$_{\rm EL,Intended}$ [$'$] & Peak$_{\rm EL,Measured}$ [$'$] & EL \%-Diff \\
\hline \\[-0.75em]
0    & 9.14$\pm$0.01  & 0.05              & 0.08              & 0.03        & 9.06$\pm$0.01  & $-$0.44            & $-$0.39            & 0.55       \\
1    & 9.35$\pm$0.01  & $-$1.87             & $-$1.79             & 0.86         & 9.33$\pm$0.02  & 4.11             & 4.12             & 0.11        \\
2    & 9.30$\pm$0.01  & 2.68              & 2.72              & 1.5         & 9.3$\pm$0.01   & 4.11             & 4.12             & 0.11        \\
3    & 9.99$\pm$0.03  & 4.61              & 4.59              & 0.22        & 9.22$\pm$0.01  & $-$0.44            & $-$0.39            & 0.54       \\
4    & 9.53$\pm$0.01  & 1.87              & 1.94              & 0.73        & 10.08$\pm$0.03 & $-$4.08            & $-$4.12            & 0.39        \\
5    & 10.31$\pm$0.03 & $-$2.68             & $-$2.72             & 1.5         & 10.39$\pm$0.03 & $-$4.08            & $-$4.12            & 0.38        \\
6    & 10.53$\pm$0.04 & $-$4.51             & $-$4.43             & 1.8         & 9.50$\pm$0.01  & $-$0.44            & $-$0.39            & 0.53      \\
\hline
\end{tabular}}
\caption{Summary of Gaussian fits to the beam response profiles shown in Figure~\ref{fig:beamMap}. Column (1): beam number; column (2): FWHM fit along XEL axis at the location of the peak response; column (3) Peak$_{\rm XEL,Intended}$ is intended location of peak response along the XEL axis; column (4) Peak$_{\rm XEL,Measured}$ is the measured location of peak response along the XEL axis; column (5): XEL \%-Offset = \textbar Peak$_{\rm XEL,Intended}$$-$Peak$_{\rm XEL,Measured}$\textbar/FWHM$_{\rm XEL}$$\times$100\%; columns (6-9): same as columns (2-5) but for EL axis.}
\label{tab:beam_fits}
\end{table*}

The response of the $i$th formed beam for each $\nu$ at each direction $\theta$ is 
\begin{equation}\label{eq:formedBeam}
I_{i}\left(\theta\right) = \left|\mathbf{w_{\rm maxSNR, i}}^H\left(\theta\right)\mathbf{a}_{i}\left(\theta)\right)\right|^2.
\end{equation}
The left panel of Figure~\ref{fig:beamMap} shows the formed beam patterns for the calibration grid around 3C295 observed for session GBT17B\_360\_04. Gaussian fits to cuts in XEL and EL at the location of each beam's peak response (red dashed lines) shown in the right hand panel and summarized in Table~\ref{tab:beam_fits} demonstrate that the FWHM of the formed beams range between approximately 9$'$ and 10.5$'$, which is comparable with the beam of the current single-pixel receiver. The offset between the measured location of the peak response of each beam and its indicated position is less than 2\% of the measured FWHM. While the outer beams are more elongated than the boresight beam, the fits to the beam profiles show deviations from a Gaussian approximation at response levels much below the FWHM. The elongated shape at levels below 10\% of the peak response is largely due to forming beams near where the sensitivity begins to drop off. For example, the elongation of the low-level response of Beam 3 corresponds to the transition from the -1 dB to -3 dB contours in the sensitivity map shown in Figure~\ref{fig:sensMap}.

%% SUBSUBSECTION ON 7-Point CALIBRATION
\subsection{7-Point Cal}\label{subsec:7PtCal}
While it is interesting to obtain detailed spatial information of the array response provided by a calibration grid, the necessary $\sim$40 minutes of total observing time (including overhead) is disadvantageous. A 7-Point calibration scan (henceforth 7Pt-Cal) can be performed in instances where telescope time is a constraint. This procedure will (1) track the area of sky minus two degrees in XEL away from the calibrator source and at the same EL offset as the center of Beams 4 and 5; (2) directly track the source (i.e. the boresight); (3) slew the telescope to put calibrator source at desired center of Beams 1-6; (4) track the area of sky minus two degrees away from the source and at the same EL offsets as the centers of Beams 1 and 2. The two reference pointings at similar EL offsets as the outer beams allow for construction of $\mathbf{R}_{\rm n}$ and also account for elevation-dependent effects, while the tracks on the desired beam centers collects the necessary response data to derive maxSNR weights. The duration of each track ranges between 10 and 30 seconds. While more efficient in terms of time than a full calibration grid, the amount of steering vectors \textbf{a}$\left(\theta\right)$ obtained during a 7Pt-Cal are only enough to set the location of the peak response for each beam and derive an SEFD. No additional information concerning the shape of the formed beams is available. This type of calibration is the primary calibration procedure for pulsar and transient observations, when detailed knowledge of the beam shape is not crucial to the science goals as compared to e.g., the overall flux sensitivity.

% SUBSECTION OUTLINING HI OBSERVATION SETUP
\subsection{\hi~Observations}\label{subsec:obsHI}
The spectral line data were collected in the fine channelized PFBCORR mode with an inherent frequency resolution of 9.47 kHz ($\sim$2 km/s at the frequency of $\hi$ emission) by steering the telescope along columns of constant longitudinal coordinates to make OTF maps. The raw dipole correlation matrices were dumped to disk every $t_{\rm int}$ = 0.5 s at angular intervals of 1.67$'$ to ensure adequate spatial Nyquist sampling; the columns/rows were spaced every 3$'$ in each DecLatMap/RaLongMap. The coordinate systems used to make our science maps include horizontal (XEL/EL), J2000, and Galactic. See Table~\ref{tab:obsSummary} and Section~\ref{subsec:HIResults} for a summary of the observational set-up for the $\hi$ sources and Sections~\ref{subsubsec:ngc6946},~\ref{subsubsec:ngc4258Field}, and ~\ref{subsubsec:galacticCenter} for the results from observations of NGC 6946, NGC 4258, and a field near the Galactic Center. 

The effective integration time of a map made with FLAG that combines all seven formed beams ($t_{\rm eff, comb}$) is derived by first computing the total effective integration time of a map made with a single beam $t_{\rm eff, map}$, multiplying this by the number of formed beams, $N_{\rm beams}$, and dividing by the map area in terms of the total number of beams contained within a map. For example, the 2$\times$2 deg$^2$ maps of NGC 6946 consists of 41 total columns ($N_{\rm columns}$), each with 72 distinct integrations ($N_{\rm int}$). Similar to the calibration procedure outlined in \citet{pingel18}, we obtain a reference spectrum from the edges of our science maps by utilizing the first and last four integrations of a particular map scan. The effective integration time for a single integration in a map from a single formed beam is therefore
%% EQUATION FOR T_INT of single scan
\begin{equation}
t_{\rm eff, int} = \frac{t_{\rm int}t_{\rm ref}}{t_{\rm int} + t_{\rm ref}} = \frac{0.5\rm~s\cdot 4\rm ~s}{0.5\rm~s + 4\rm~s} = 0.444 \rm~s;
\end{equation}
$t_{\rm eff, map}$ then follows from $N_{\rm rows}\times N_{\rm int}\times t_{\rm eff, int}=$1312 s and increases to $t_{\rm eff, map}\times N_{\rm beams} = 1312 \times 7 = 9184$ sec for combined map. The FWHM of the approximately Gaussian boresight beam is 9.1$'$, which corresponds to an angular area of 1.1331$\times\left(9.1'\right)^2\sim$ 0.026 deg$^{2}$. The area in terms of the number of beams is then 4 deg$^{2}$/0.026 deg$^{2}$ $\sim$153 beams. The final $t_{\rm eff, comb}$ is then just $t_{\rm eff, comb}$ = 9184 s / 153 beams $\sim$ 60 s/beam. These $t_{\rm eff, comb}$ values are listed listed in the Notes column of Table~\ref{tab:obsSummary} for each science map and can be used in the ideal radiometer equation to calculate the theoretical noise value in the final images.

\section{Data Reduction}\label{sec:dataReduction}
The data reduction and calibration of the $\hi$ data was performed with a custom Python software packages {\tt pyFLAG}\footnote{\url{https://github.com/nipingel/pyFLAG}}. This section summarizes the scripts available to perform the post-correlation beamforming, flux calibration, and imaging of FLAG spectral line data. 

%% SUBSECTION ON FILLING SOFTWARE
\subsection{Post-Correlation Beamforming}\label{subsec:postcorr_beamforming}

A scan performed with FLAG produces several types of ancillary FITS\footnote{\url{https://fits.gsfc.nasa.gov/standard40/fits_standard40aa-le.pdf}} files that contain important metadata such as the antenna positions and LO settings. These metadata must be collated and combined with the raw covariances stored in FITS files to create a single dish FITS (SDFITS\footnote{\url{https://safe.nrao.edu/wiki/bin/view/Main/SdfitsDetails}}) file that can be manipulated in GBTIDL\footnote{\url{http://gbtidl.nrao.edu/}}, just as data from the single-pixel receiver. The unique format of the raw FLAG dipole covariances necessitate custom {\tt pyFLAG} software to collate all the metadata and perform the post-correlation beamforming (i.e., Equation~\ref{eq:beamformedPower}) to generate an SDFITS file for each formed beam that contains beamformed spectra in units of raw counts. 

This software suite contains all the necessary Python and GBTIDL code with which to calibrate and image spectral line data from FLAG. In both correlator modes (i.e., PFBCORR and CALCORR), each GPU runs two correlator threads making use of the xGPU library\footnote{\url{https://github.com/GPU-correlators/xGPU/tree/master/src}}, which is optimized to work on FLAG system parameters. Each correlator thread handles 1/20th of the total bandwidth made up of either 25 \textit{non-contiguous} coarse frequency channels with 303.18 MHz resolution or 160 contiguous fine channels with 9.47 kHz resolution and writes the raw output to disk in a FITS file format. The data acquisition software used to save these data to disk was borrowed from development code based for the Versatile GBT Astronomical Spectrometer (VEGAS) engineering FITS format. The output FITS file from each correlator thread is considered a `bank' with a unique X-engine ID (XID; i.e., the correlator thread) ranging from 0 to 19 that is stored in the primary header of the FITS binary table; there are therefore 20 distinct FITS files created for each scan. Reading and sorting the covariances stored within each bank FITS file --- whether placing the non-contiguous 25$\times$20 coarse channels in the correct order or stitching together the the 160$\times$20 contiguous fine channels --- is crucial a step within the {\tt pyFLAG} software.

%% FIGURE SHOWING COVARIANCE MATRIX STRUCTURE
\begin{figure*}
\centering
\includegraphics[width =5.5in]{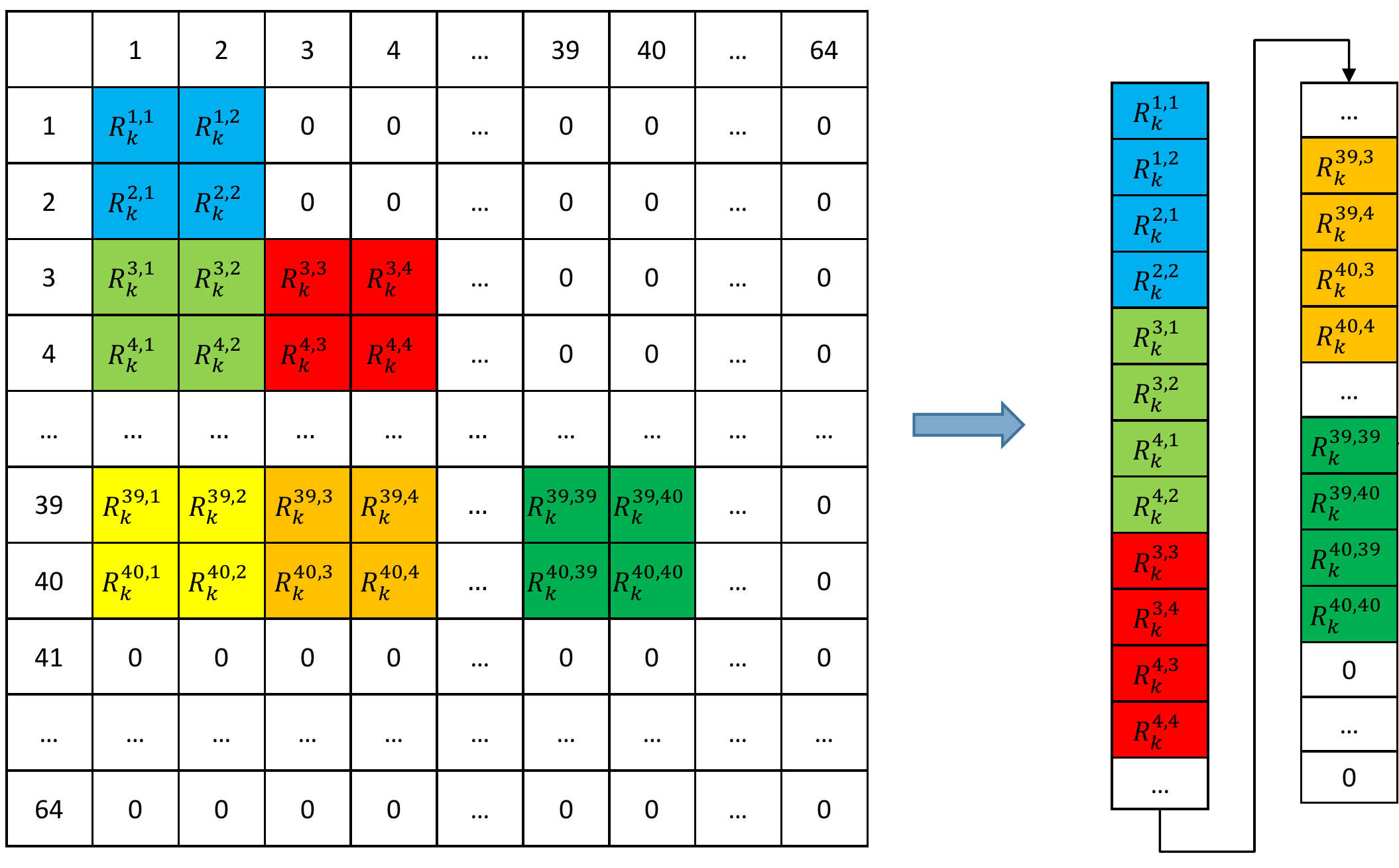}
\caption{The structure of a covariance matrix used in beamforming. The numbers preceding each row/column correspond to the dipole element. Each element of the matrix stores the covariance between dipole elements $i$ and $j$ for a single frequency channel, $k$. The output is ordered in a flattened one-dimensional array that needs to be reshaped into a 40$\times$40 matrix before beamforming weights can be applied. Additionally, due to xGPU limitations, the output size is 64$\times$64, which results in many zeros that need to be thrown away in data processing.}
\label{fig:covarianceMatrixExample}
\end{figure*}

The raw data output for both CALCORR and PFBCORR correlator modes are the covariance matrices containing the covariance between individual dipole elements. However, due to xGPU limitations, the covariance matrices are shaped 64$\times$64 and flattened to a one-dimensional (1D) data vector.

An example of how the covariance values are ordered is illustrated in Figure~\ref{fig:covarianceMatrixExample}. Here, $R_{\rm k}^{\rm i,j}$ corresponds to the covariance between dipole $i$ and $j$ at frequency channel $k$. Most of the transpose pairs (e.g., $R_{\rm k}^{\rm 1,3}$) are shown as zero because they are not included in the 1D data array that is saved to disk in order to preserve disk space. Additionally, since only the covariances between the first 40 data streams (19 dipoles$\times$2 polarizations$+$ 2 spare channels), there is a large portion of zeros appended onto the end of the 1D data array. The correlator output represents the block lower triangular portion of the large 64x64 covariance matrix shown in this figure sorted in row-major order, where a block corresponds to a colored 2x2 sub-matrix. The reduction scripts treat each 4-element contiguous chunk of the 1D data vector as a block and place it into the larger covariance matrix in row-major order. Once the first 40 rows have been filled in, a conjugate transpose operation is performed to fill in the missing covariance pairs and the remaining zeros are discarded.

When in CALCORR mode, the bank file corresponding to XID ID 0 contains covariances matrices for frequency channels 0 to 4, 100 to 104, ..., 400 to 404; the XID 1 bank file stores covariance matrices for frequency channels, 5 to 9, 105 to 109, ..., 405 to 409. However in PFBCORR mode, the covariance matrices for channels 0 to 159 are stored in the bank file corresponding to XID 0 and continue in a contiguous fashion such that the bank file corresponding to XID 19 stores data for frequency channels 3039 to 3199. The logic during data reduction is to process each frequency channel individually, then sort the result into the final bandpass based on the XID and mode in which the data were taken. The scripts that drive the creation of an SDFITS file are {\tt PAF\_Filler.py} --- in essence the `main' function of the program --- and the two modules {\tt metaDataModule}.py and {\tt beamformerModule.py}.

The foremost step in the filling and calibration process of FLAG data is to solve Equation~\ref{eq:eigen} for the dominant eigenvector using the R$_{\rm s}$ and R$_{\rm n}$ covariance matrices obtained from calibration scans to determine the maxSNR complex beamforming weights. This is performed with the {\tt pyFLAG} python script, {\tt pyWeights.py}, which also generates a series of 20 FITS files (one for each bank). Each beamformer weight FITS file contains a binary table consisting of 14$\times$3200 elements: (7 beams$\times$2 polarizations)$\times$(64 elements$\times$25 frequency channels$\times$2 for the complex pair). The headers of these FITS files also contain the beam offsets (in arcminutes), calibration set filenames, beamforming algorithm, and XID.

Once the weights have been generated, {\tt PAF\_Filler.py} can be run. This script reads in and unpacks each bank FITS file to pass the raw data 1D covariances to the beamformer object created by {\tt beamformingModule.py}. Each bank FITS files is processed in parallel to maximize efficiency. 

Within this module, the FITS files storing the complex beamforming weights are read in and organized into the form of a 2D numpy array of complex data-type, with the rows representing the 25 coarse frequency channels and columns represent the correlations of the `40' dipoles (19$\times$2 dual polarization dipoles plus 2 spare data channels). Once the complex weights are in the correct format, the raw 1D covariances recorded for each integration are reordered and transposed according to the block row-major scheme summarized in Figure~\ref{fig:covarianceMatrixExample} and reshaped into a 3D {\tt numpy} array of complex data-type with rows and columns both representing the correlations between dipoles and the third axis representing a given frequency channel. %More explicitly, for each plane of this data cube, 
%the first element of the first column holds the auto-correlation ($\rho$) between the first dipole (dp) $\rho_{dp1,dp1}$, the second element is the cross-correlation between dipoles 1 and 2, or $\rho_{dp1,dp2}$, the third holds $\rho_{dp1,dp3}$, ..., and the last holds $\rho_{dp1,dp40}$. 
%Since the correlations are redundant, the corresponding transposed element in a row is simply the conjugate value of the column value. 
The final returned cube for each integration has dimensions of 40$\times$40$\times\mathbf{N_{\rm chan}}$, where $N_{\rm chan}$ is again number of frequency channels per bank file --- either 25 or 160, depending on whether FLAG is operating in CALCORR or PFBCORR mode, respectively. Note two important aspects: (1) the irrelevant correlations caused by xGPU limitations are thrown away at this stage; (2) some rows and columns contain zeros as they correspond to two unused data streams. Equation~\ref{eq:beamformedPower} is then applied to each plane of the correlation cube to construct a beamformed bandpass in units of raw counts. A 2D array containing the beamformed bandpass for each integration is returned to {\tt PAF\_Filler.py} and sorted into global data buffers. The software will recognize the mode based on the number of channels stored in a bank FITS file. When in PFBCORR mode, where 100 coarse channels are sent through a PFB implementation to obtain a total of 3200 fine channels, the beamformer weight for an input coarse channel will be applied across the 32 corresponding output fine channels.

After each bank FITS file for a particular scan is processed, the filled global data buffers are passed to a metadata object created by {\tt metaDataModule.py}. This object collates all associated metadata, applies the beam offsets to the recorded antenna positions, and perform the necessary Doppler corrections to the topocentric frequencies. Once all corrections to the spatial and spectral coordinates have been made, the binary FITS tables are combined and appended to a primary Header Data Unit and returned to {\tt PAF\_Filler.py} where the final SDFITS file is written to disk. The process then repeats until all beams for all observed objects are processed. Comprehensive documentation and usage examples are available at \url{https://github.com/nipingel/pyFLAG}.

%% SUBSECTION ON HI CALIBRATION/IMAGING
\subsection{Spectral Line Calibration and Imaging}\label{subsec:HIReduction}

%% TABLE SUMMARIZING INDIVIDUAL SESSION TSYS/ETA and SYSTEM EQUIVALEN FLUX DENSITIES FOR SCIENCE SCANS
\begin{table*}
\centering
\resizebox{\textwidth}{!}
{\begin{tabular}{lccccc}
\hline \hline
\\[-1.0em]
Session & Beam & Scan Type & SEFD [Jy] & Calibration Source & Calibration Source Flux [Jy] \\
\hline \\[-0.75em]
%% AGBT16B_400_12
\textbf{GBT16B\_400\_12 (NGC 6946)} & & & & \\
 & 0 & Grid & 14$\pm$1 & 3C295 & 22.15 \\
 & 1 & Grid & 15$\pm$2 & 3C295 & 22.15 \\
 & 2 & Grid & 15$\pm$2 & 3C295 & 22.15 \\
 & 3 & Grid & 15$\pm$1 & 3C295 & 22.15 \\
 & 4 & Grid & 16$\pm$1 & 3C295 & 22.15 \\
 & 5 & Grid & 16$\pm$2 & 3C295 & 22.15 \\
 & 6 & Grid & 17$\pm$2 & 3C295 & 22.15 \\
%% AGBT16B_400_13
\textbf{GBT16B\_400\_13 (NGC 6946)} & & & & \\
 & 0 & Grid & 14$\pm$1 & 3C123 & 22.15 \\
 & 1 & Grid & 15$\pm$2 & 3C123 & 22.15 \\
 & 2 & Grid & 15$\pm$2 & 3C123 & 22.15 \\
 & 3 & Grid & 15$\pm$1 & 3C123 & 22.15 \\
 & 4 & Grid & 16$\pm$1 & 3C123 & 22.15 \\
 & 5 & Grid & 16$\pm$2 & 3C123 & 22.15 \\
 & 6 & Grid & 17$\pm$2 & 3C123 & 22.15 \\
%% AGBT17B_360_03
\textbf{GBT17B\_360\_03 (NGC 4258 Field)} & & & & \\
 & 0 & Grid & 16.4$\pm$0.3 & 3C295 & 22.15 \\
 & 1 & Grid & 17.0$\pm$0.4 & 3C295 & 22.15 \\
 & 2 & Grid & 16.2$\pm$0.6 & 3C295 & 22.15 \\
 & 3 & Grid & 16.9$\pm$0.6 & 3C295 & 22.15 \\
 & 4 & Grid & 18.1$\pm$0.4 & 3C295 & 22.15 \\
 & 5 & Grid & 17.4$\pm$0.4 & 3C295 & 22.15 \\
 & 6 & Grid & 17.5$\pm$0.4 & 3C295 & 22.15 \\
%% AGBT17B_360_04
\textbf{GBT17B\_360\_04 (NGC4258 Field)}* & & & & \\
 & 0 & Grid & 9.3$\pm$0.2 & 3C295 & 22.15 \\
 & 1 & Grid & 9.5$\pm$0.3 & 3C295 & 22.15 \\
 & 2 & Grid & 9.6$\pm$0.2 & 3C295 & 22.15 \\
 & 3 & Grid & 9.6$\pm$0.3 & 3C295 & 22.15 \\
 & 4 & Grid & 9.7$\pm$0.3 & 3C295 & 22.15 \\
 & 5 & Grid & 9.5$\pm$0.3 & 3C295 & 22.15 \\
 & 6 & Grid & 9.7$\pm$0.2 & 3C295 & 22.15 \\
%% AGBT17B_455_01
\textbf{GBT17B\_455\_01 (G353$-$4.0)} & & & & \\
 & 0 & 7Pt-Cal{\dag} & 10$\pm$2 & 3C348 & 48.14 \\
 & 1 & 7Pt-Cal & 10$\pm$2 & 3C348 & 48.14 \\
 & 2 & 7Pt-Cal & 10$\pm$2 & 3C348 & 48.14 \\
 & 3 & 7Pt-Cal & 10$\pm$1 & 3C348 & 48.14 \\
 & 4 & 7Pt-Cal & 10$\pm$1 & 3C348 & 48.14 \\
 & 5 & 7Pt-Cal & 10$\pm$3 & 3C348 & 48.14 \\
 & 6 & 7Pt-Cal & 10$\pm$3 & 3C348 & 48.14 \\
\hline
\end{tabular}}
\caption{Summary of derived system properties in XX Polarization from calibration scans used to make the science images; {\dag} denotes that $\nu_{0}$ was set to 1450.00000 MHz for Beams 0-6; {\ddag} denotes that $\nu_{0}$ was set to 1450.8484 MHz.}
\label{tab:SEFDSummary} 
\end{table*}

After post-correlation beamforming to obtain spectra in units of raw system counts, flux calibration of $\hi$ data can begin. We calculate the SEFD (see Equation~\ref{eq:SEFD} and discussion in Section~\ref{subsec:calGrid}) from the CALCORR calibration scans. The flux measured on the sky is
%% EQUATION FOR SKY FLUX
\begin{equation}\label{eq:Ssrc}
S_{\rm sky} = {\rm SEFD}\left(\frac{P_{\rm On}}{P_{\rm Off}} - 1\right).
\end{equation}
As discussed above, we obtain a reference spectrum to use as $P_{\rm Off}$ from the edges of our science maps by taking the mean power in each frequency channel for the first and last four integrations of a particular map scan. $P_{\rm On}$ in Equation~\ref{eq:Ssrc} is then the raw power in each integration recorded during the scan. The SEFD values used to scale the raw power ratios for each beam and each session are computed with Equation~\ref{eq:SEFD} as discussed in Section~\ref{subsec:calGrid} and summarized in Table~\ref{tab:SEFDSummary}. The flux calibration scripts are written in GBTIDL and driven with a python wrapper, {\tt PAF\_edgeoffkeep\_parallel.py}, in order to calibrate each of the seven beams in parallel. The mean SEFD over all beams included in our science maps is 12.3$\pm$0.3 Jy/beam. However, this value is biased by measurements taken before improvements in calibration procedures (see the discussion below); a more typical value after improvements is 9.8$\pm$0.4 Jy. If we assume an $\eta$ of 0.65 \citep{boothroyd2011} for the sake of direct comparison with the single-pixel receiver, and use the more characteristic SEFD value of 10 Jy, Equation~\ref{eq:SEFD} gives a $T_{\rm sys}$ of 18.5 K. While this assumption of $\eta$ does not consider specific parameters of the FLAG receiver, such as the large spillover from the illumination pattern of individual dipoles and their mutual coupling, this $T_{\rm sys}$ value is consistent with both the existing single-pixel receiver ($\sim$ 18 K) and the $T_{\rm sys}$/$\eta$ measurements of \citet{roshi18} at 1.4 GHz ($\sim$25-35 K). The overall sensitivity of FLAG is discussed in Section~\ref{subsec:TsysResults}.

The measured $T_{\rm sys}$/$\eta$ is directly related to the SEFD (i.e., Equation~\ref{eq:Ta_eta}). Consistent SEFD values are critical for accurately reproducing measurements of flux on the sky between observing sessions and making comparisons between the data collected by FLAG and other instruments. We see that the overall SEFD values progressively converge to the single-pixel value and observe a consistent decrease in the variation between beams and session-to-session with subsequent observing runs. We attribute the steady reduction in both measured SEFD values and associated scatter to consistent improvements to the calibration strategies used to obtain and maintain bit and byte-lock --- such as the introduction of scripts to automate this process. We stress that our most accurate flux measurements are obtained from our later observing sessions, specifically GBT17B\_360\_04 and beyond. We therefore note that the maps presented in Section~\ref{subsec:HIResults} from previous sessions are done so with the caveat that the overall flux scale has high uncertainty relative to later sessions. Furthermore, since the overall flux scale of an OTF spectral map depends on both the area of the assumed telescope beam and the width of the convolution function used to interpolate the individual samples to a regular image grid \citep{mangum2007}, we present $\hi$ flux density profiles only from sessions where the weights were derived from a full calibration grid to ensure the beam response is fully characterized over the FoV. 

%% FIGURE SHOWING SCALLOPPING OF A RAW BEAMFORMED SPECTRA
\begin{figure}
%\centering
\includegraphics[width = \columnwidth]{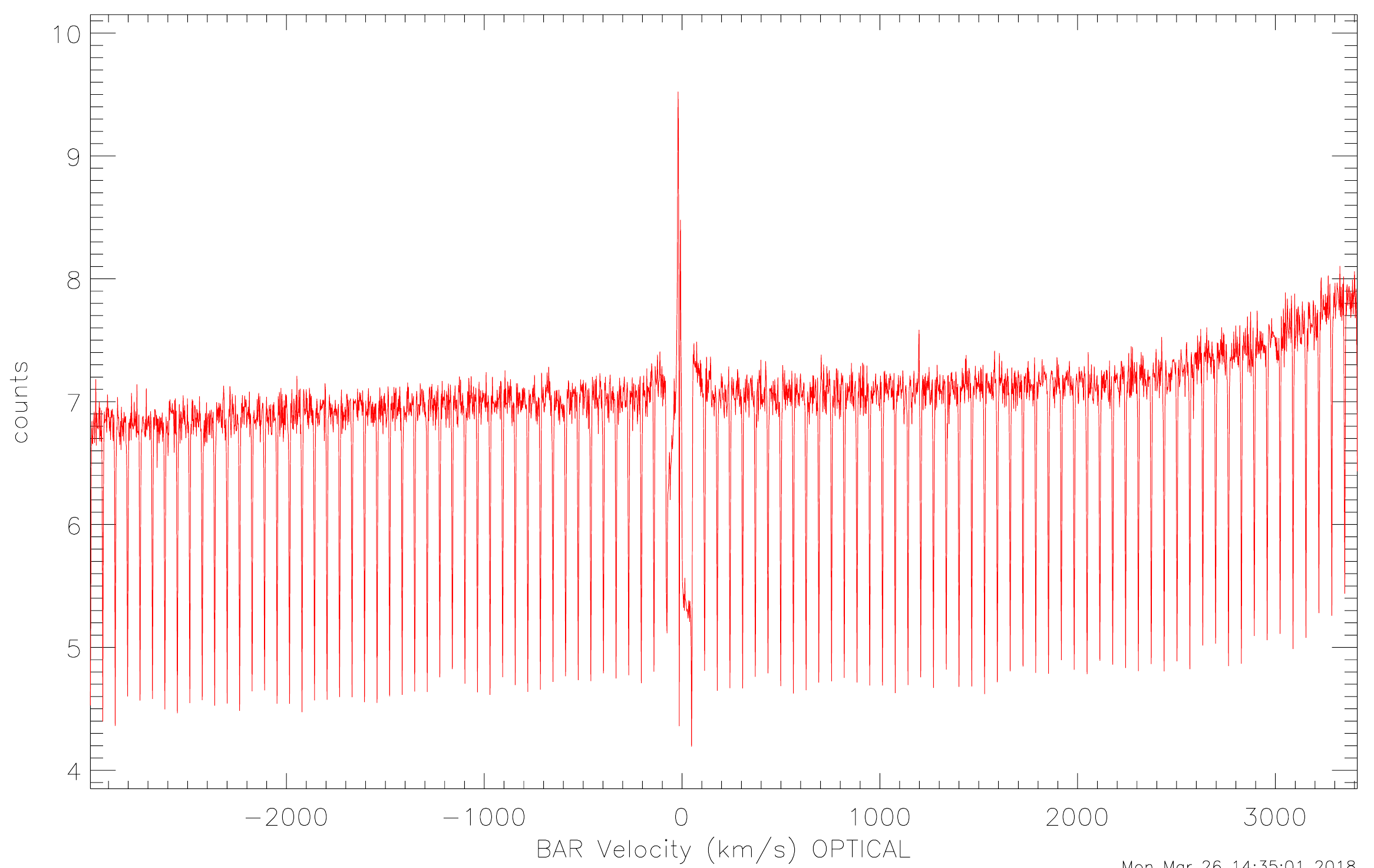}
\includegraphics[width = \columnwidth]{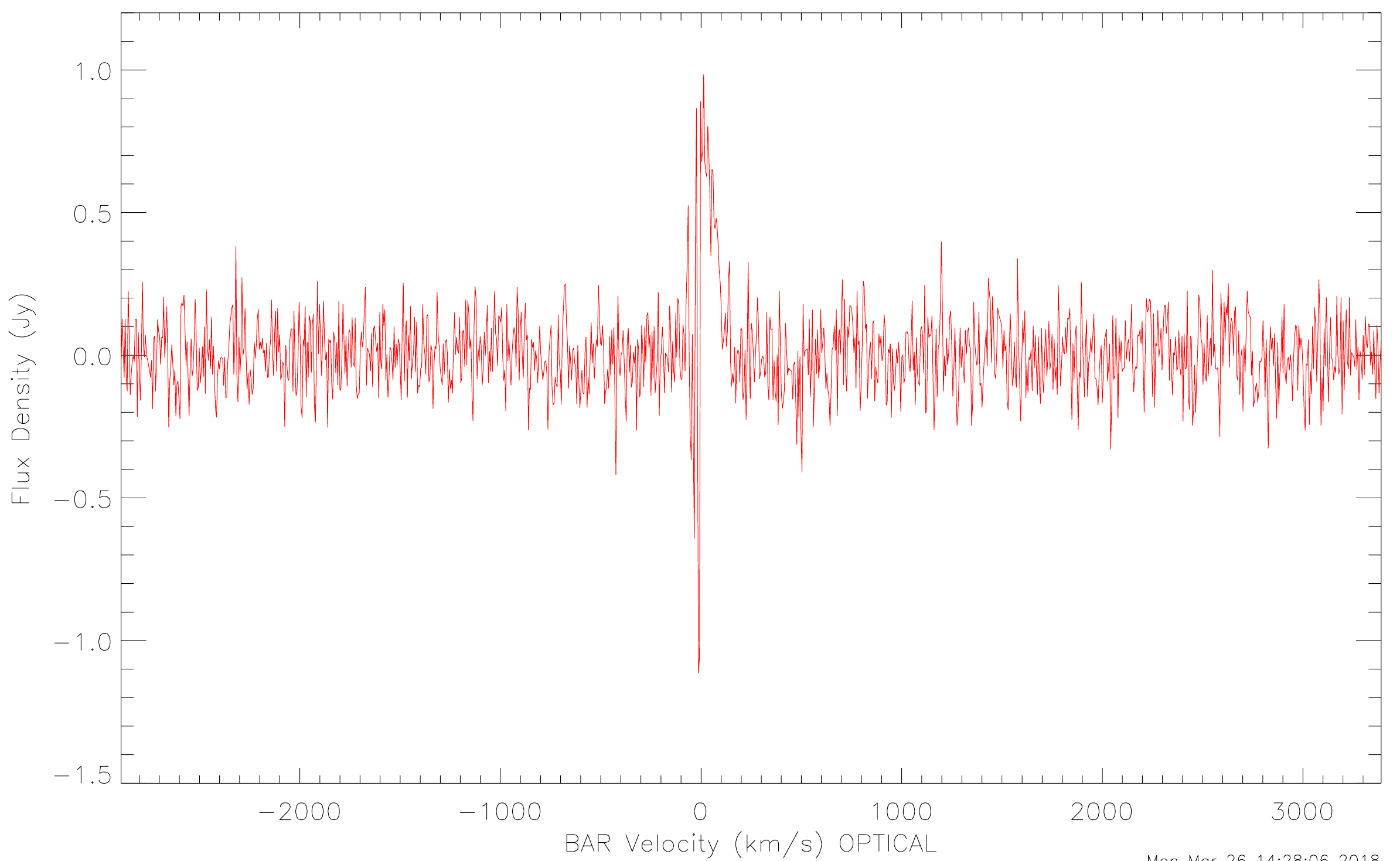}
\caption{An example of an uncalibrated, beamformed spectrum taken from the 35th integration of the 19th column of a DecLatMap scan of NGC6946. The $-$3 dB drop in power (i.e., `scalloping') is an artifact of the two step PFB implementation of the back end (see text). \textit{Bottom}: The calibrated version of the above spectrum. While the scalloping behavior appears to be mitigated, the signal aliasing at the edge of a coarse channel is still present.}
\label{fig:exampleSpectra}
\end{figure}

\subsection{Bandpass Scalloping}\label{subsec:scalloping}
An example of a raw and calibrated integration when the system is in PFBCORR mode is shown in Figure~\ref{fig:exampleSpectra}. The nulls, or `scalloping', seen every 303.18 kHz (every 32 fine frequency channels) in the top panel is a consequence of the two stage PFB architecture approach currently implemented in the back end. As the raw complex time series data are processed within the ROACHs, a response filter is implemented in the coarse PFB such that the adjacent channels overlap at the $-$3 dB point to reduce spectral leakage (power leaking in from adjacent channels). However, this underlying structure becomes readily apparent after the fine PFB implemented in the GPUs. The scalloping therefore traces the structure of each coarse channel across the bandpass. While the structure is somewhat mitigated in the calibrated data (since there is a division by a reference spectrum), power variations caused by spectral leakage in the transition bands of the coarse-channel bandpass filter result in residual structure. Additionally, this scheme leads to signal aliasing stemming from the overlap in coarse channels. Such near-coarse-channel-band-edge aliasing artifacts are present in a number of other existing astronomical two-stage zoom spectrometers. These artifacts do not hinder the performance of FLAG in terms of sensitivity, but a fix for the signal aliasing is a priority going forward. A provisional fix with the capability to provide both coarse and narrowband spectra is realized by a two-stage channelizer architecture. The first implemented in the ROACH and the second as part of PFBCORR mode in the GPU. Both stages of processing use PFBs for computationally efficient channelization. In our case we are implementing critically sampled PFBs at both stages. To remove spectral artifacts (aliasing, scalloping) the first stage channelizer must be an oversampled PFB to allow adjacent channels to overlap. In the output of the second stage critically sampled PFB (PFBCORR), the overlapped region is discarded to eliminate artifacts.

%A proposed implementation will double \textbf{the number of input coarse channels, as well as implementing two 256-point Fast Fourier Transforms (FFT) that run at different rates of 303 kHz and 606 kHz. The output is then interleaved to avoid a $-$3 dB drop off, at the penalty of slightly reducing the usable bandwidth from 151.18 MHz to 121.2 MHz due to the discarding of additional channels.}

The scalloping can be completely mitigated by dithering the frequency such that a subsequent map has a central frequency that is either 151.59 kHz (or one-half of a coarse channel width) above or below the initial central frequency setting. Because the scalloping is caused by overlap of the input 100 coarse channels into the PFB, there are 98 instances of drops in power across the total 3200 fine channels, with each dip corresponding to 56 kHz or six fine channels corresponding to the three channels at each edge of a coarse channel. The channels affected by the scalloping are known beforehand and do not change regardless of LO setting. In a dithered observation, the affected channels from observations at both frequency settings can be blanked with the {\tt chanBlank\_parallel.py} script before imaging to ensure no signal is aliased in the final maps. These blanked calibrated spectra are smoothed with a Gaussian kernel to a final resolution of 5.2 km s$^{-1}$ and imaged with the {\tt gbtgridder}\footnote{\url{https://github.com/GreenBankObservatory/gbtgridder}} tool, utilizing a Gaussian-tapered circular Bessel gridding function. Note that we present images of only the XX linear polarization due to complications with the YY polarization signal chain during our two observing runs that has since been rectified. We account for the use of a single polarization in our calculations of sensitivity and comparison to equivalent single-pixel data. 

%% SECTION SUMMARIZING RESULTS
\section{Results}\label{sec:results}

%% SUBSECTION ON BEAMFORMER
\subsection{Beamformer Weights}

%% FIGURE SHOWING VARIATION OF WEIGHT PHASE/AMP
\begin{figure}
\includegraphics[width = \columnwidth]{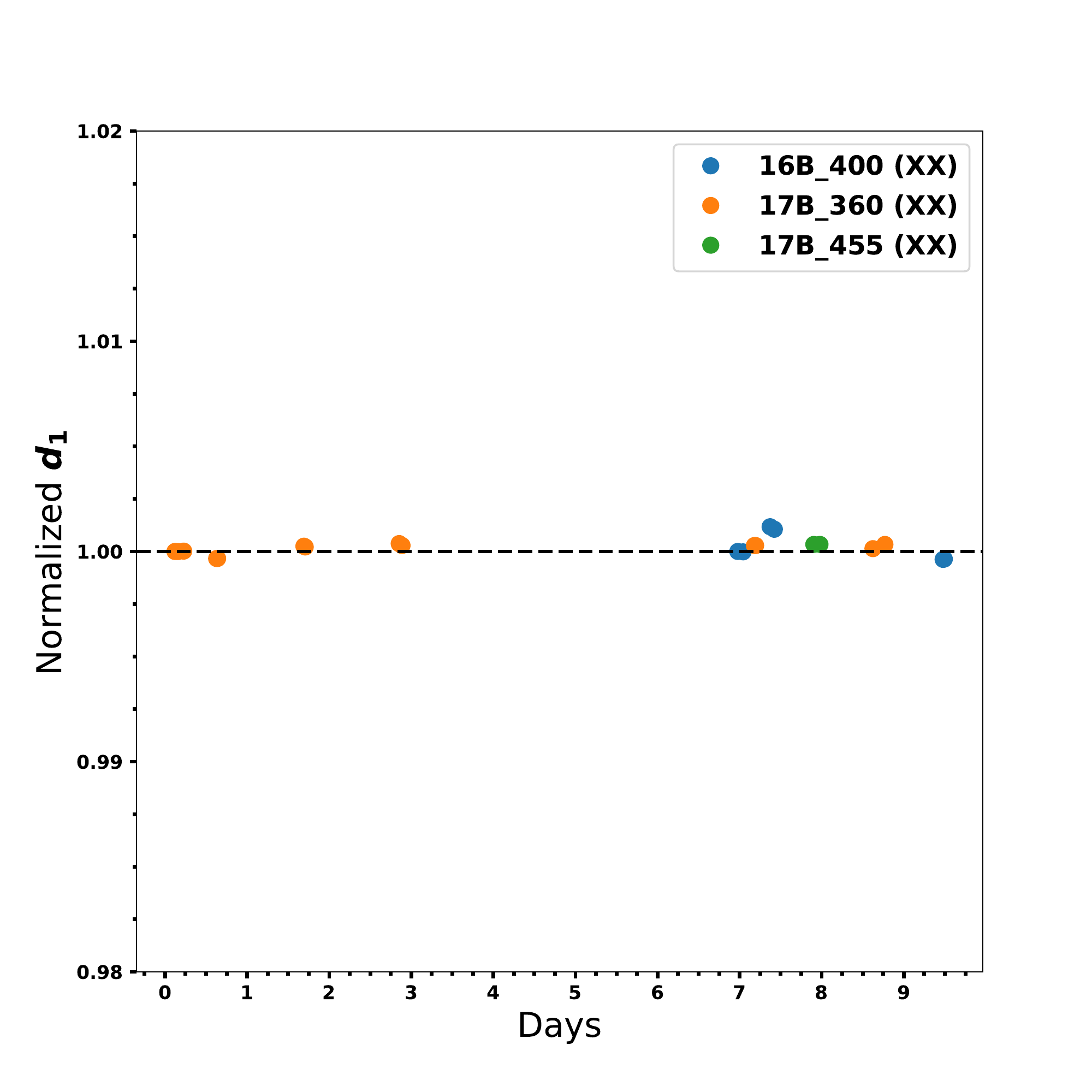}
\caption{Variation of the phase distance metric between subsequent beamforming weights for the boresight beam as a function of time. The $d_{\rm 1}$ values are corrected for the bulk phase offset according to Equation~\ref{eq:bulkPhaseOffset} and normalized by the first $d_{\rm 1}$ value from each observing epoch for clarity.}
\label{fig:weightPhaseAndAmp}
\end{figure}

The calibration procedure described in Section~\ref{subsec:calGrid} contributes to $\sim$40 minutes of overhead and, in principle, can remain valid for several weeks if bit/byte lock is not lost. However, since lock is currently lost with every change in the local oscillator setting, it is recommended that an observer derive fresh beam weights at the beginning of each observing session. Other reasons re-calibration may be necessary include: large variations in the contribution of spillover and sky noise to the signal model and the relative electronic gain drift between dipoles \citep{jeffs2008}. Important factors that impact the quality of the weights are robust bit and byte locks, constraining the desired steering vector for a formed beam, and utilizing a sufficiently bright calibration source to adequately characterize the system response when on and off source.

While the current state of the FLAG system effectively requires new beamforming weights every session, it is still interesting to explore how the complex weight vectors derived from a given calibration observation vary with time. Studying the variations will help reveal characteristic properties and behavior of the weights that demonstrate the stability of the system with time.   

Recall that a given element in the weight vector is a complex number that contains the amplitude and phase information to be applied to the output of a given dipole in order to steer a beam in the desired direction. Beam steering is primarily influenced by varying the amplitude of the weights applied to each dipole. Given the reliable placement of our formed beams demonstrated in Figure~\ref{fig:beamMap}, we wish to investigate how the formed beam responses are influenced by the second-order effect of phase variations. To measure the difference in phase, a distance metric can be defined
%% EQUATIONS DESCRIBING DISTANCE MEASUREMENTS TO MEASURE VARIATIONS IN BEAMFORMER WEIGHT QUALITY
\begin{equation}\label{eq:phaseDistanceMetric}
d_1 = \lvert\lvert \mathbf{a_1} - \mathbf{\tilde{a}_2} \rvert\rvert, 
\end{equation}
where $\mathbf{a_1}$ and $\mathbf{a_2}$ are the vector norms (i.e., the square root of the sum of each element's squared complex modulus) of the weight vectors, or $\mathbf{w_1}/\lvert\lvert \mathbf{w_1}\rvert\rvert$ and $\mathbf{w_2}/\lvert\lvert \mathbf{w_2}\rvert\rvert$, respectively. The vector $\mathbf{\tilde{a}_2}$ represents the subsequent weight vector that has been corrected for the bulk phase offset between the two vectors. This bulk phase offset arises from the steering vectors, which are found by solving for the dominant eigenvector in the generalized eigenvalue problem in Equation~\ref{eq:eigen}. Since eigenvectors are basis vectors that have arbitrary scaling, it is the unknown scaling of the phase between calibration data sets that contributes to the bulk phase offset. A subsequent weight vector can be phase aligned to some initial weight vector by first making the first element of $\mathbf{a_1}$ real and then computing 
%%EQUATION THAT SOLVES FOR BULK PHASE OFFSET
\begin{equation}\label{eq:bulkPhaseOffset}
\hat{\phi} = \angle\left(\mathbf{a^H_2} \mathbf{a_1}\right),
\end{equation}
where $\hat{\phi}$ is the angle of the product of $\mathbf{a^H_2} \mathbf{a_1}$. The correction for the bulk phase offset is therefore a complex scaling factor applied to all phases in the latter weight vector to ensure the phase differences in the remaining dipoles arise strictly from the systematic (e.g., bit/byte lock) and instrumental effects between different weight calibrations. The phase aligned weight vector is therefore $\mathbf{\tilde{a}_2} = e^{i\hat{\phi}}\mathbf{a_2}$.

Because the distance metric $d{_1}$ is the overall magnitude of an element-wise difference between two $M$ element vectors, it encapsulates all the phase differences between respective dipoles into a single quantity. Small variations in $d_1$ over time indicate similar phases (save for the bulk phase offset due, in part, to new bit/byte lock) between the derived weight vectors, meaning the directional response of the array is stable over the time span of a typical observing run; thus, the beam pattern shape remains relatively unchanged. 

%The two vectors used to compute $d_1$ are unit normalized and thus does not describe the scale variations between two weight vectors. Large variations in the relative scale will affect the gain stability over the field-of-view and cause difficulties in flux calibrations between observing sessions. To detail this behavior a separate distance metric can be defined
%% EQUATION FOR SCALE DISTANCE METRIC
%\begin{equation}\label{eq:scaleDistanceMetric}
%d_2 = \lvert\vert \mathbf{w_2} \rvert\rvert/ \lvert\vert \mathbf{w_1} \rvert\rvert. 
%\end{equation}

Figure~\ref{fig:weightPhaseAndAmp} shows the variation of the normalized $d_1$ distance metric as a function of time for the boresight beam. We see similar trends for each of the outer beams and no discernible difference between types of calibration scans performed. The initial set of beamformer weights (i.e., $\mathbf{a_1}$) is taken to be the first set of weights derived for that particular observing run. We compare all subsequent weights from a given observing run with this first set. The time values are taken to be the difference between the mean Modified Julian Date (MJD) values associated with a given calibration scan, with the initial MJD taken to be from the first calibration scan in a given observing run. The scatter in the phase variations is well below the 1\% level, indicating that the directional response to identical coincident signals is very similar over time, which ensures the peak response is  reliably located in the desired direction on the sky and similar beam structure between sessions.   

\begin{figure*}
\centering
\includegraphics[width = 3.5 in]{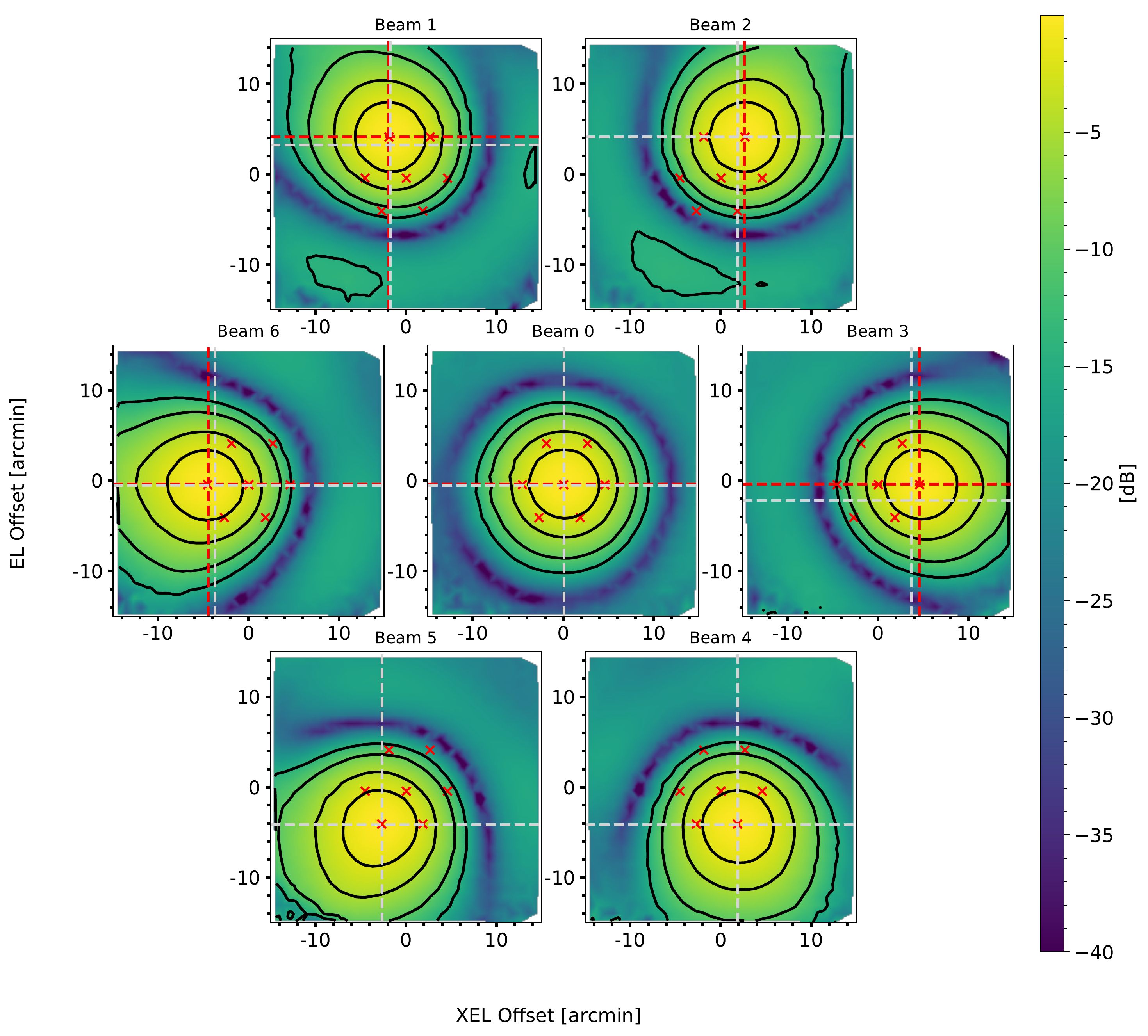}
\includegraphics[width = 3.5 in]{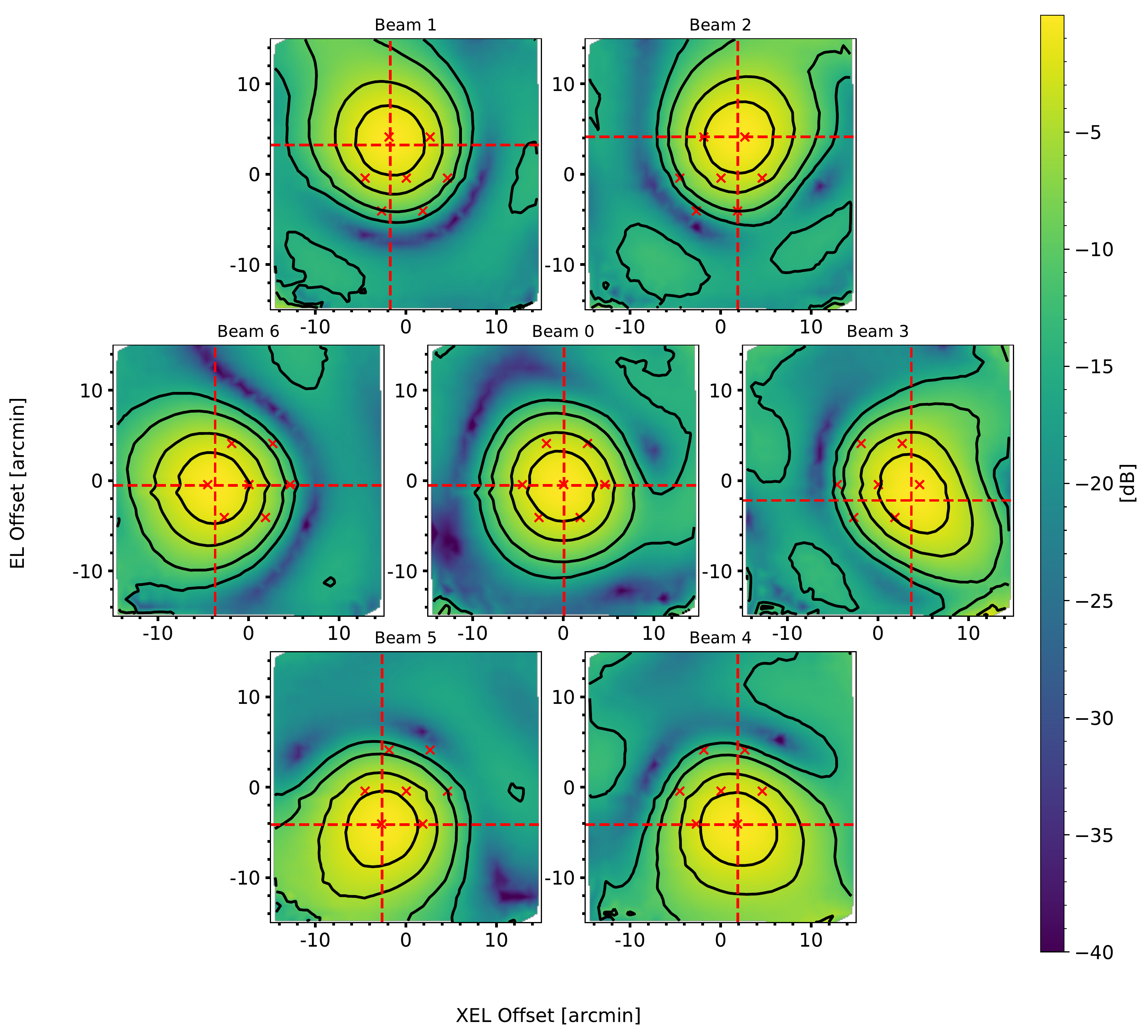}
\includegraphics[width = 3.5 in]{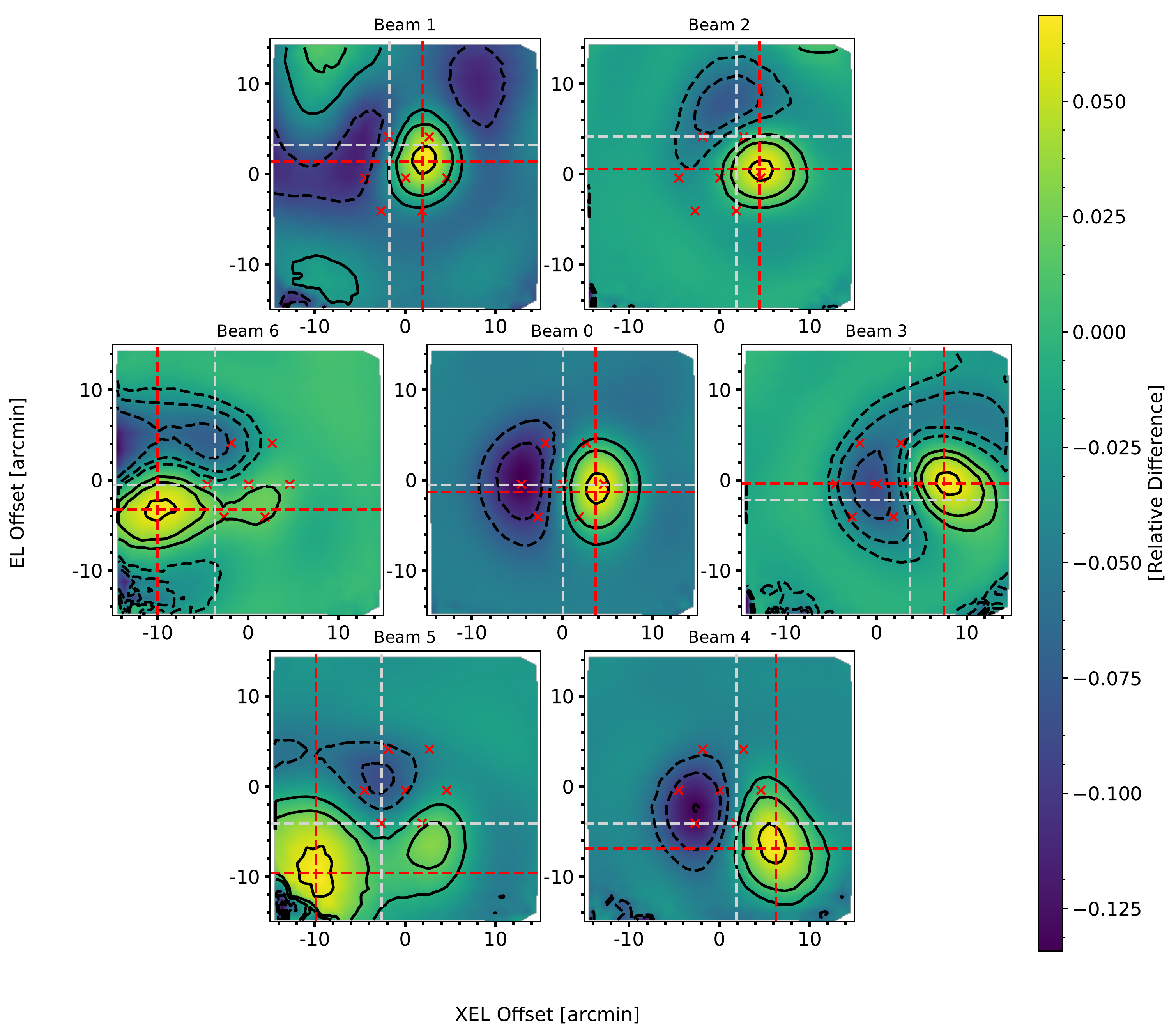}
\includegraphics[width = 3.5 in]{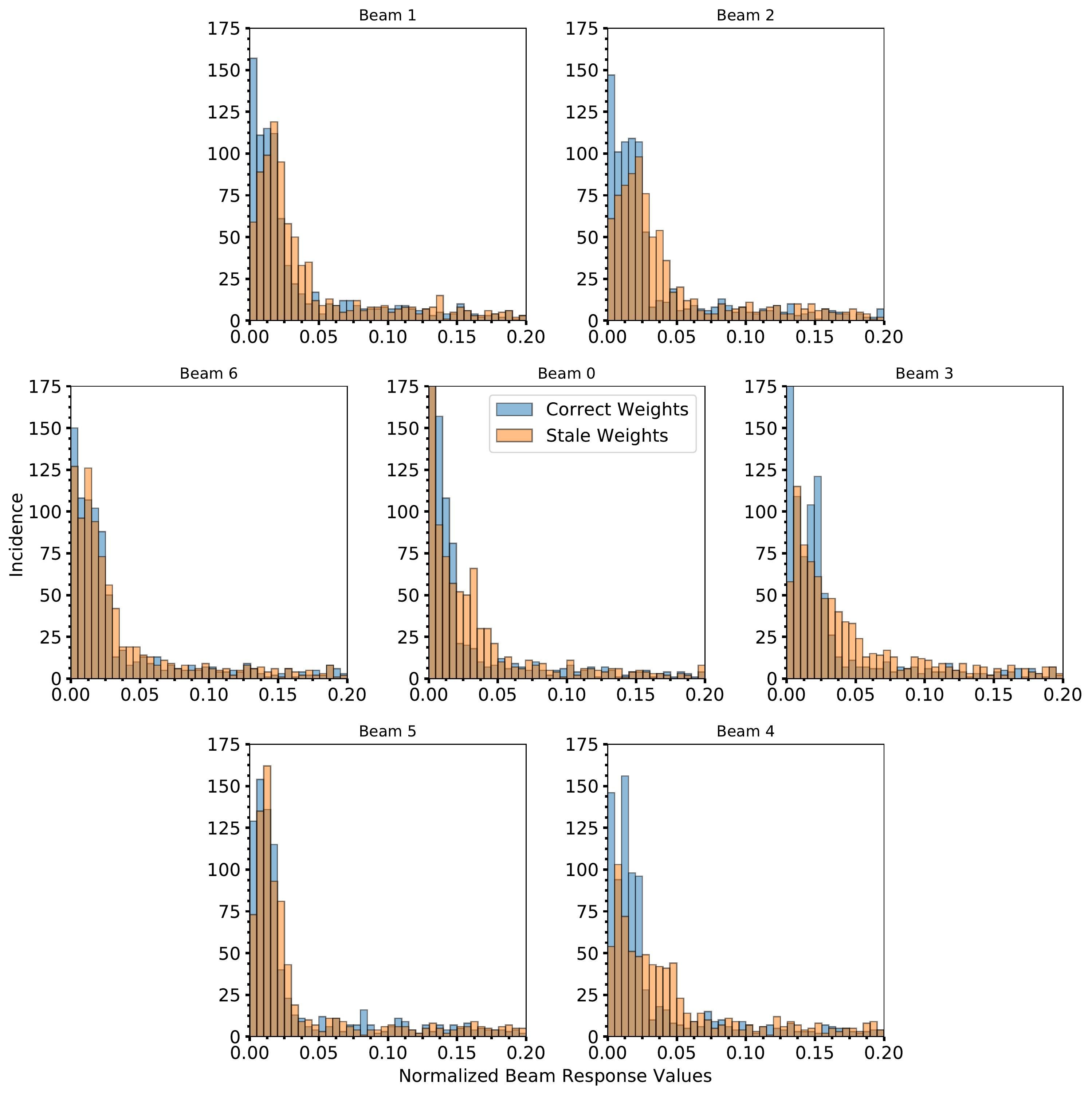}
\caption{Formed beam patterns and resulting histograms wherein beamforming weights derived from the calibration grids performed during the GBT17B\_360\_04 and GBT17B\_360\_07 observing sessions were applied to steering vectors from the GBT17B\_360\_04 calibration grid. The contours, red dash lines, and $\times$ symbols are the same as in Figure~\ref{fig:beamMap}. The white vertical and horizontal dashed lines correspond to the red lines from the upper right panel as a reference to the shift in peak response caused by the application of stale weights.
\textit{Top left}: beam pattern derived using the correct weights. \textit{Top right}: beam pattern derived using stale (i.e., from GBT17B\_360\_07) weights.
\textit{Bottom left}: the difference of the top left and right panels. The solid (dashed) contours denote the 90\%, 50\%, and 25\% level of the maximum relative difference between each formed beam.
\textit{Bottom right}: Partial histograms of the beam response values shown in the upper panels. The range of response values is chosen to highlight the difference at the levels of the sidelobes.}
\label{fig:subsequentWeights}
\end{figure*}

Figure~\ref{fig:subsequentWeights} demonstrates the effect of varying beamforming weights on the measured beam shapes. Weights derived from the calibration grids performed during the GBT17B\_360\_04 observing sessions were applied to steering vectors from the GBT17B\_360\_07 calibration grid. Weights derived during an earlier session applied to steering vectors from a subsequent session are considered to be stale, since the sample delays required to achieve a previous bit/byte lock will produce a different phase response. By examining the changes in overall beam shape, the locations of the peak response of each formed beam relative to the desired pointing center, and change in sensitivity (i.e., Equation~\ref{eq:sens}), we are able to investigate the stability of these beam weights between observing sessions. 

The beams formed with stale weights retain their overall Gaussian shape. While the peak response of the stale boresight beam is close to the desired pointing center, the peak responses of some of the outer beams, specifically Beam 3, shifts significantly. The difference map in the bottom left panel reveals that the relative phase offset in the stale weights degrades the sidelobe structure, shifting the low-level beam response towards the edge of the FoV. The change in the low-level beam response is further illustrated in the partial histograms shown in the bottom left panel. The peaks in the histograms that represent the sidelobe structure shift to higher values and become broader, indicating a change in the overall beam shape below the 10\% level. The change in shape of each distribution is due to the relative phase offset present in the stale weights. 

We compute a measure of sensitivity for each formed beam using Equation~\ref{eq:sens}. The value of $\mathbf{w^{\rm H}}\mathbf{R_{\rm s}}\mathbf{w}$ is taken to be the maximum power value at 1420 MHz in a 7-Pt calibration scan when the calibrator is centered in a given beam, and the $\mathbf{w^{\rm H}}\mathbf{R_{\rm n}}\mathbf{w}$ value is the average power value in the nearest reference pointing at that same frequency. Taking the ratio of the sensitivity values between beams formed from stale weights to those formed with the correct weights reveals an average drop of 56\% between all formed beams. Overall, the beams formed with stale weights are stable above the 50\% level of the peak response. However, the application of stale weights results in beam patterns that are, on average, half as sensitive and possess altered directional responses to the same incident signal at the levels of the first sidelobes. An observer should account for the overhead to perform at least a 7Pt-Cal to derive contemporaneous weights. 

A calibration strategy deemed `word lock'  that, in principle, will allow observers to reuse previously derived weights is nearing deployment. This procedure accounts for the variable amount of sample delays between each bit/byte lock cycle by utilizing the time-shift property of the Fourier Transform to insert shifts in the full 16-bit/2-byte word. By inserting the optimal amount of shifts that minimizes the variation in phase across frequency relative to a reference dipole \citep{burnett2017masters}, the phase response of the previous set of weights will now apply to the current state of the system.

%% FIGURE SHOWING VARIATION OF WEIGHT PHASE/AMP

%% SUBSECTION ON TSYS/ETA
\subsection{Sensitivity as a Function of Frequency}\label{subsec:TsysResults}

%% FIGURE SHOWING FLUX OFFSET IN N6946 DATA FROM AGBT16B_400_12
\begin{figure*}
\centering
\includegraphics[width = 7 in]{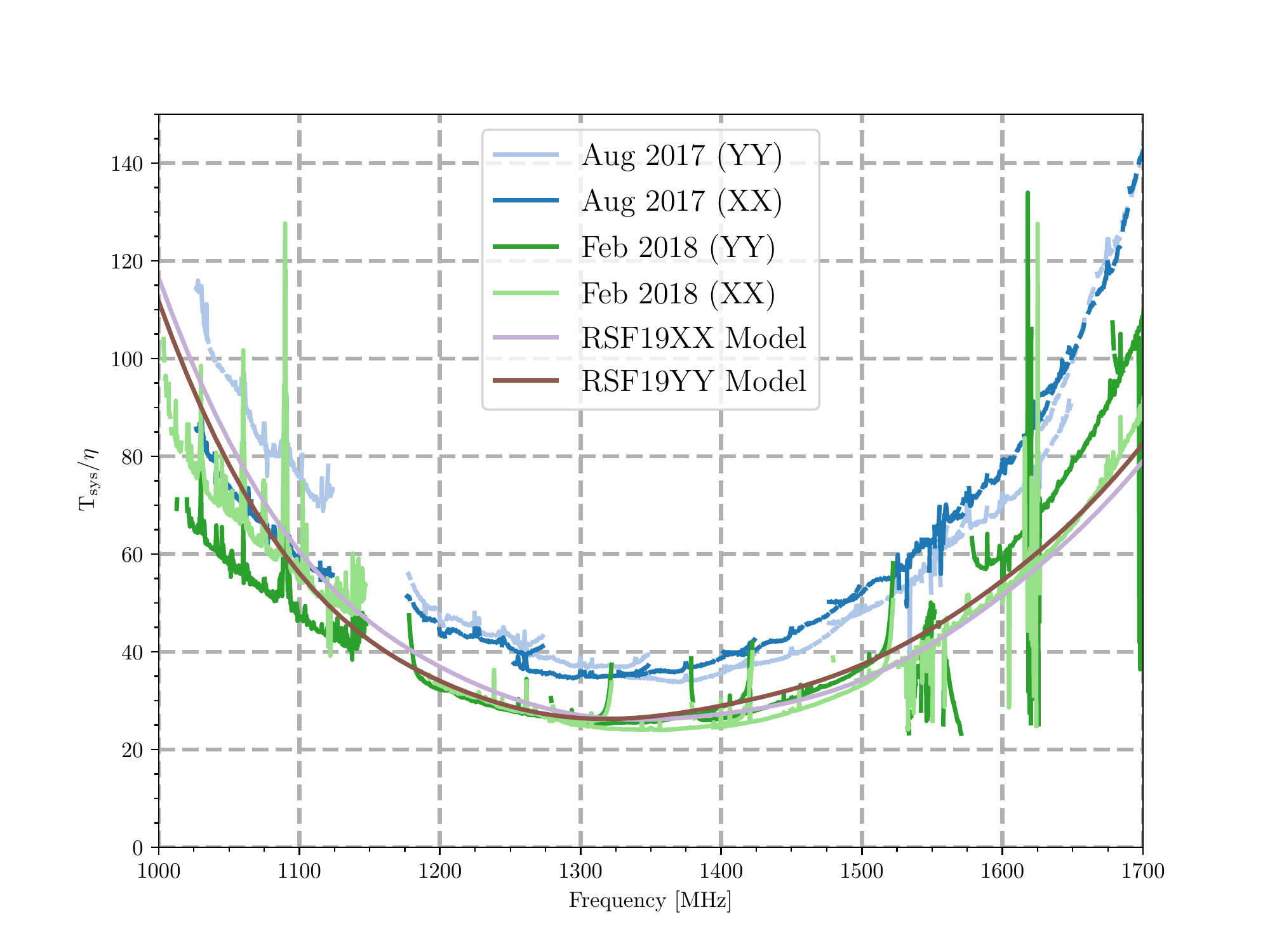}
\caption{\label{fig:TsysOverlay} $T_{\rm sys}$/$\eta$ (see Equation~\ref{eq:Tsys_eta}) as a function of frequency derived for a set of 7Pt-Cal scans in which the LO was sequentially shifted by 50 MHz. The PAF model results form \citet{roshi2019_Models} corresponding tot he two polarizations are marked RSF19XX and RSF19YY. These model results correspond to a thermal transition length of 9.1 cm and its loss of 1 K. See \citet{roshi2019_Models} for further details.}  
\end{figure*}

Figure~\ref{fig:TsysOverlay} shows the result of Equation~\ref{eq:Tsys_eta} derived from frequency sweep observations performed as engineering tests for several of the commissioning runs. The goal of this test is to characterize the sensitivity over a wide range of frequencies and identify frequencies most affected by narrowband RFI features. We performed a series of 7Pt-Cal scans with the LO set to 50 MHz increments beginning at 1100 MHz and continuing up to 1700 MHz. For each calibration scan, we calculate $T_{\rm sys}$/$\eta$ as a function of the 150 MHz bandpass for each formed beam at the coarse channel resolution of 0.30318 MHz and merge the results. 

As can be expected with \textbf{significant} improvements to the system made between subsequent commissioning runs, $T_{\rm sys}$/$\eta$ decreases as a function of epoch for both polarizations with the February 2018 calibration data showing the lowest observed $T_{\rm sys}$/$\eta$. Since $T_{\rm sys}$/$\eta$ and SEFD depend on one another, we also attribute this trend to improvements made to the signal processing algorithms of the back end and calibration strategies to obtain bit/byte lock. Specifically, a correction to increase the digital gain to avoid a bit underflow when the data in the ROACH is reduced to 8-bit/8-bit real and imaginary values just before packetization was implemented for the February 2018 observing runs.

The measured $T_{\rm sys}$/$\eta$ are compared to several PAF system models (see Figure~\ref{fig:TsysOverlay}). In general, these models are produced by first obtaining the modified full polarization response pattern of the individual dipole elements embedded in the array. Finite element solutions of electromagnetic equations were used to obtain these response patterns. The patterns along with a model of the GBT optics were used to predict the full polarized electromagnetic field pattern in the antenna aperture and to characterize ground spillover. These results were used
to compute the signal covariance and the noise covariance due to the ground spillover and sky background. A noise model for the cryogenic LNAs is utilized to pre dict the receiver contribution to the noise covariances. The maxSNR beamforming algorithm is then applied to the signal and noise covariances to predict the final $T_{\rm sys}$/$\eta$ at a given frequency. See \citet{roshi2019_Models} (hereafter RSF19) for further details on modeling.

The measured $T_{\rm sys}$/$\eta$ for the February observing run are largely consistent with the models at a frequency of 1.4 GHz. Overall, the measured sensitivity across functional frequency range of the receiver are consistent with expectations models with only moderate narrowband RFI near the $\hi$ transition. The discrepancy between the models and measurements at lower frequencies may be due to differences between the modeled and actual roll-off of the analog filter. Obvious RFI artifacts present between 1000 MHz and 1100 MHz and near 1625 MHz need to be considered when planning potential observations of radio recombination lines and the OH 1665 MHz and 1667 MHz transitions. While we only show results for the boresight beam, the trends are similar for all outer beams. 

\subsection{\hi~Results}\label{subsec:HIResults}

\subsubsection{NGC 6946}\label{subsubsec:ngc6946}
The external galaxy, NGC 6946, was chosen as the first science target for $\hi$ on the basis of ample GBT single-pixel data available for comparison (e.g., \citealt{pisano14}). The presence of high-velocity gas from galactic fountain activity \citep{boomsma08} and an $\hi$ filament, possibly related to recent accretion \citep{pisano14}, and several smaller nearby companions are also ideal features to test the sensitivity of this new receiver.  

\begin{figure*}
\includegraphics[width = \textwidth]{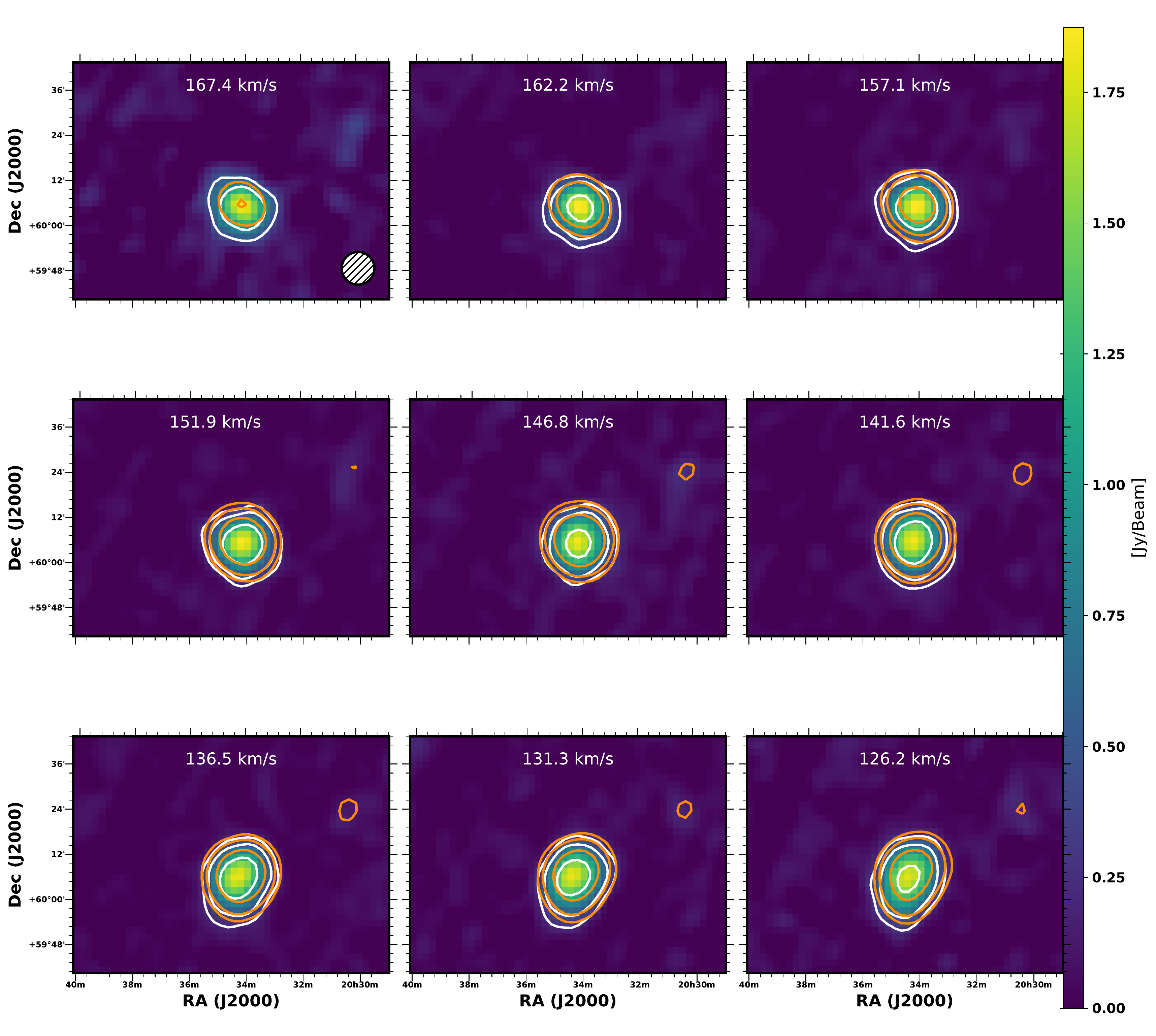}
\caption{\label{fig:N6946ChanMaps} Channel maps of NGC 6946 and nearby companions. $\hi$ emission detected by the FLAG boresight beam is represented by the color scale and white contours, while emission detected by the single-pixel receiver is denoted by orange contours. Both sets of contours begin at the 130 mJy/Beam level ($\sim$3$\sigma_{\rm meas}$ in Table~\ref{tab:summaryOfNoiseAndSS}) and continue at 10 and 25 times that level.}
\end{figure*}

\begin{figure}
\includegraphics[width = \columnwidth]{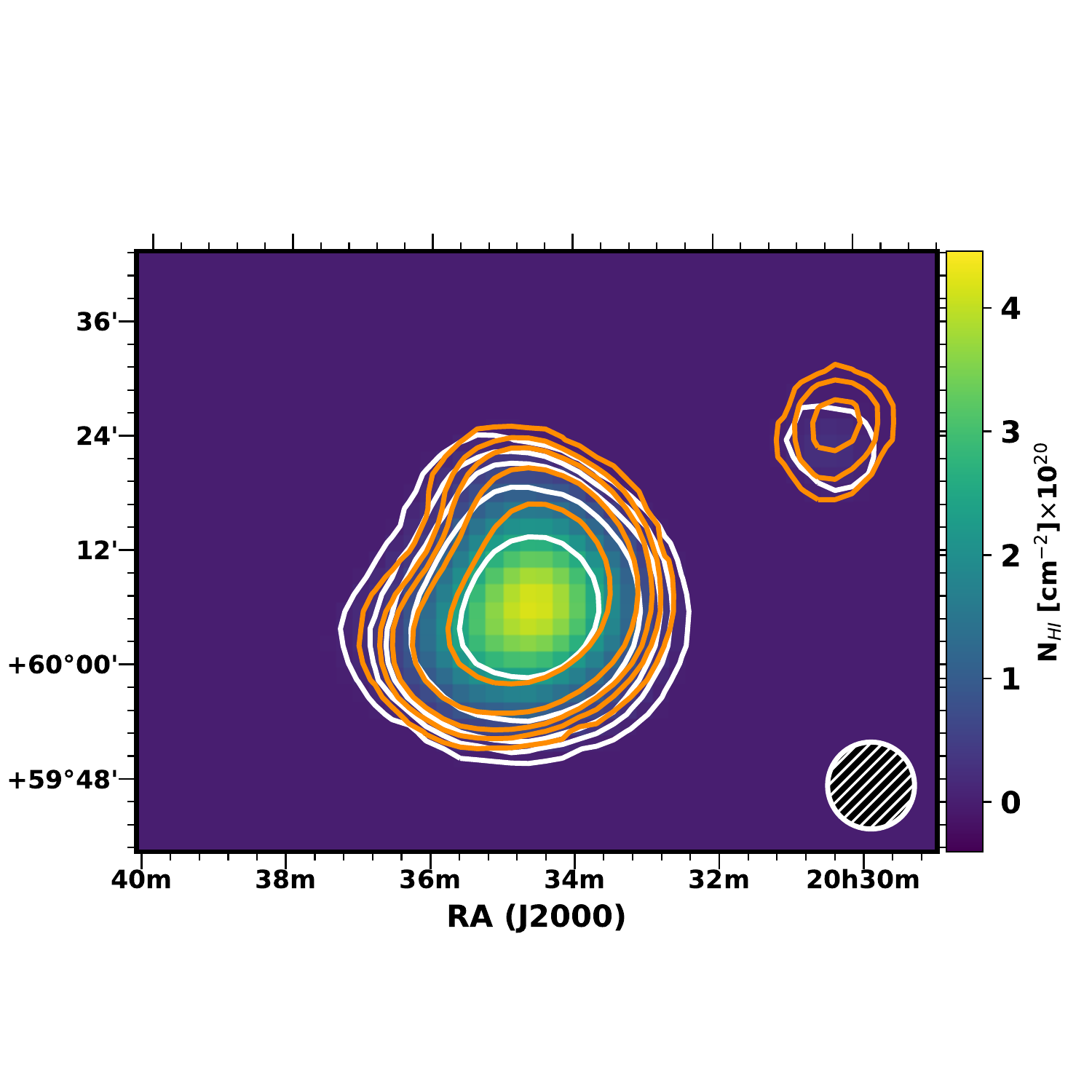}
\caption{\label{fig:N6946Mom0Map} $\hi$ column density map of NGC 6946. FLAG data is represented by the color scale and white contours, while the single-pixel equivalent column density levels are overlaid in orange. The outer contour is at a level of 1$\times$10$^{19}$ cm$^{-2}$, which represents a 3$\sigma$ detection over the integrated 11.4 km s$^{-1}$ to 181.4 km s$^{-1}$ velocity range, while the inner contours go as 5, 10, and 25 times that level. We have assumed the emission is optically thin and a similar gain of 1.86 K/Jy to convert the FLAG data to units of brightness temperature.}
\end{figure}

This source was observed in the horizontal celestial coordinate system to ensure beam offsets, which are determined in the same coordinate frame by definition, were correct. The images presented here are 2$^{\circ}\times$2$^{\circ}$ large and had the central frequency ($\nu_{0}$) set to 1450.0000 MHz in the topocentric Doppler reference frame. For a direct comparison with previous single-pixel data and a single FLAG beam, we show channel maps from the boresight beam in Figure~\ref{fig:N6946ChanMaps} to demonstrate that FLAG effectively reproduces single-pixel observations. Overall, the FLAG and single-pixel contours agree well with the slight offsets in the lowest level contours are attributed to the fact that the FLAG data are a factor of almost 10 times less sensitive than the single-pixel data. The difference in sensitivity between these two maps also explains the non-detection of the two unresolved companions, UGC11583 and L149, in the channel maps from a single beam. Figure~\ref{fig:N6946Mom0Map} reveals the presence of the two companions once data from all seven FLAG beams are combined in a single map. Here, the slight differences at the lowest contour levels likely arise from the complicated beamshape and sidelobe structure resulting from averaging the seven distinct formed beams. 

%% FIGURE SHOWING CHANNEL MAPS OF NGC 6946 FIELD

%% FIGURE SHOWING MOM0 MAP OF NGC6946

\subsubsection{NGC 4258 Field}\label{subsubsec:ngc4258Field}
NGC 4258 resides in the Canes Venatici II Group \citep{deVauc75}, which is comprised of several companions including the late-type galaxies NGC 4288 and UGC 7408 to the southwest and J121811.04+465501.1 --- a low surface brightness dwarf galaxy \citep{Lian07} --- slightly to the southeast. The most appealing target in the field is a prominent $\hi$ filament extending from NGC 4288 that points towards NGC 4258. This filament was seen previously with the 76.2m Lovell telescope at the
Jodrell Bank Observatory (UK) as part of the $\hi$ Jodrell All Sky Survey (HIJASS); \citep{wolfinger13}. It is classified as an `$\hi$ cloud' with the designation HIJASS J1219+46; no known optical counterparts are observed over the spatial extent of the $\hi$ emission. The single-pixel data used as a comparison, which was collected during a GBT survey to provide the single-dish counterpart to the high-resolution Westorbork Radio Synthesis Telescope (WSRT) Hydrogen Accretion in LOcal GAlaxieS (HALOGAS) Survey \citep{heald11, pingel18}, shows a peak flux of $\sim$0.06 Jy and projected physical scale of $\sim$80 kpc, assuming the same distance as to NGC 4258. 

A total of six 1.5$^{\circ}\times$2$^{\circ}$ maps evenly split over two separate observing sessions were performed. Improvements to how the beam offsets were applied in the custom reduction software enabled mapping in equatorial coordinates. The first session the $\nu_{0}$ set to 1450.0000 MHz and 1449.84841 MHz, respectively, to circumvent the frequency scallopping (see Figure~\ref{fig:exampleSpectra} and discussion in Section~\ref{subsec:HIReduction}). The relative weak flux, extended nature, and complex kinematics originating from a possible tidal interaction between HIJASS J1219+46 and other group members provide an excellent benchmark for the mapping capabilities of FLAG.

\begin{figure*}
\includegraphics[width = \textwidth]{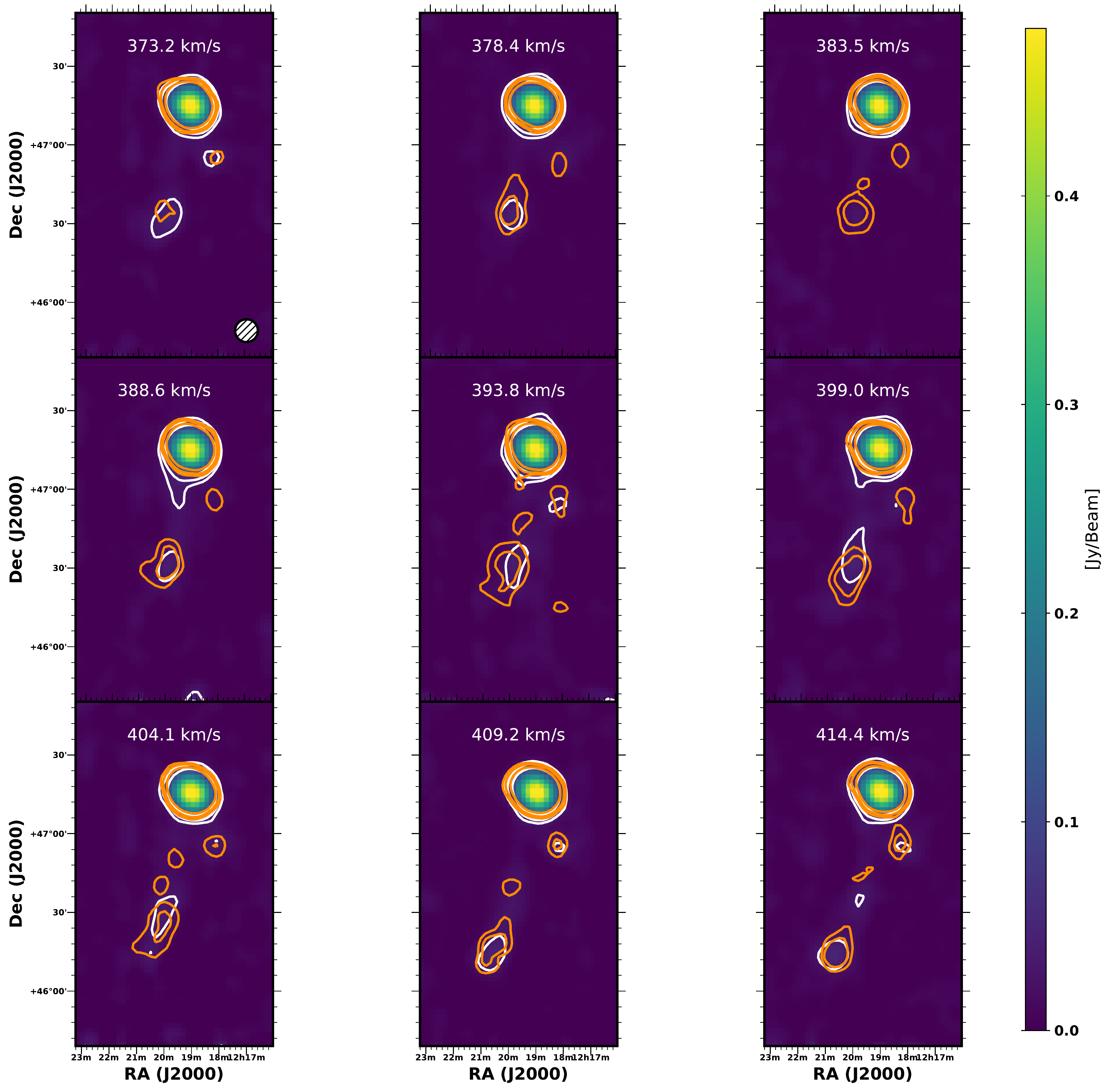}
\caption{\label{fig:N4258ChanMaps} Channel maps of the NGC 4258 Field. $\hi$ emission detected by FLAG is represented by the color scale and white contours, while emission detected by the single-pixel receiver is denoted by orange contours. Both sets of contours begin at the 27 mJy/Beam level ($\sim$3$\sigma_{\rm meas}$ in Table~\ref{tab:summaryOfNoiseAndSS}) and continue at 5 and 10 times that level.}
\end{figure*}

\begin{figure}
\includegraphics[width = \columnwidth]{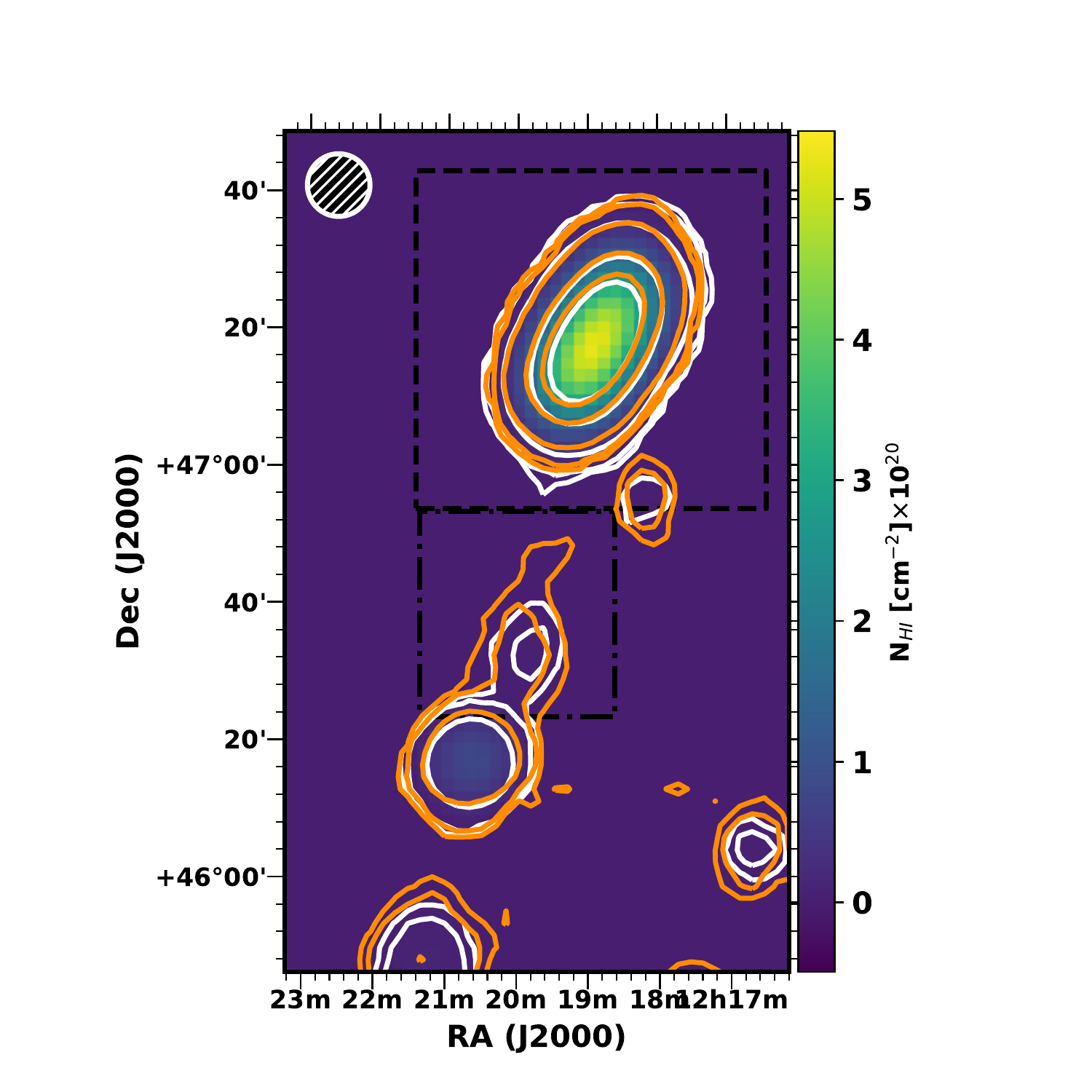}
\caption{\label{fig:N4258MomentMap} 
$\hi$ column density map of the NGC4258 Field observed by FLAG (color scale and white contours) with equivalent single-pixel data (orange contours) overlaid. The contour levels beginning at 2$\times$10$^{18}$ cm$^{-2}$, which represents a 5$\sigma$ detection over a 20 km s$^{-1}$ line, and continuing at 15, 100, and 500 and 1000 times that level. The dashed and dot-dashed rectangles denote the angular areas over which the flux profiles shown in Figure were integrated.}  
\end{figure}

The channel maps of the NGC 4258 Field in Figure~\ref{fig:N4258ChanMaps} contain data from all seven beams from sessions 17B\_360\_03 and 17B\_360\_04, with data from the former session being scaled by the factor listed in Table~\ref{tab:summaryOfNoiseAndSS} to ensure a consistent flux scale. While there is not a specific cause for the relatively large scale factor of $\sim$0.6 between observing sessions, we again note that scripts to automate the bit/byte locking procedure were used for the first time before the latter session, which has shown to significantly increase the stability of the system over the course of multiple observing sessions that use the same LO configuration. These channel maps demonstrate that FLAG can reproduce the features of diffuse structures detected by the single-pixel receiver when mapping at similar sensitivities. The contours tracing the filament, HIJASS J1219+46, extending from NGC 4288 between the velocities of 378 km s$^{-1}$ and 409 km s$^{-1}$ are in agreement, albeit for the lowest level contours that are affected by the unconstrained sidelobe levels. The $\hi$ column density image in Figure~\ref{fig:N4258MomentMap}, in which a mask was applied such that only pixels with a S/N$>$3 are included in the final image shows very good correspondence at all contour levels between the FLAG and single-pixel data with a clear detection of the low-level emission associated with the $\hi$ cloud. 

%% FIGURE SHOWING CHANNEL MAPS OF NGC4258 FIELD

%% FIGURE SHOWING MOM0/MOM1 MAPS OF NGC4258 FIELD

\begin{figure*}
\includegraphics[width = \textwidth]{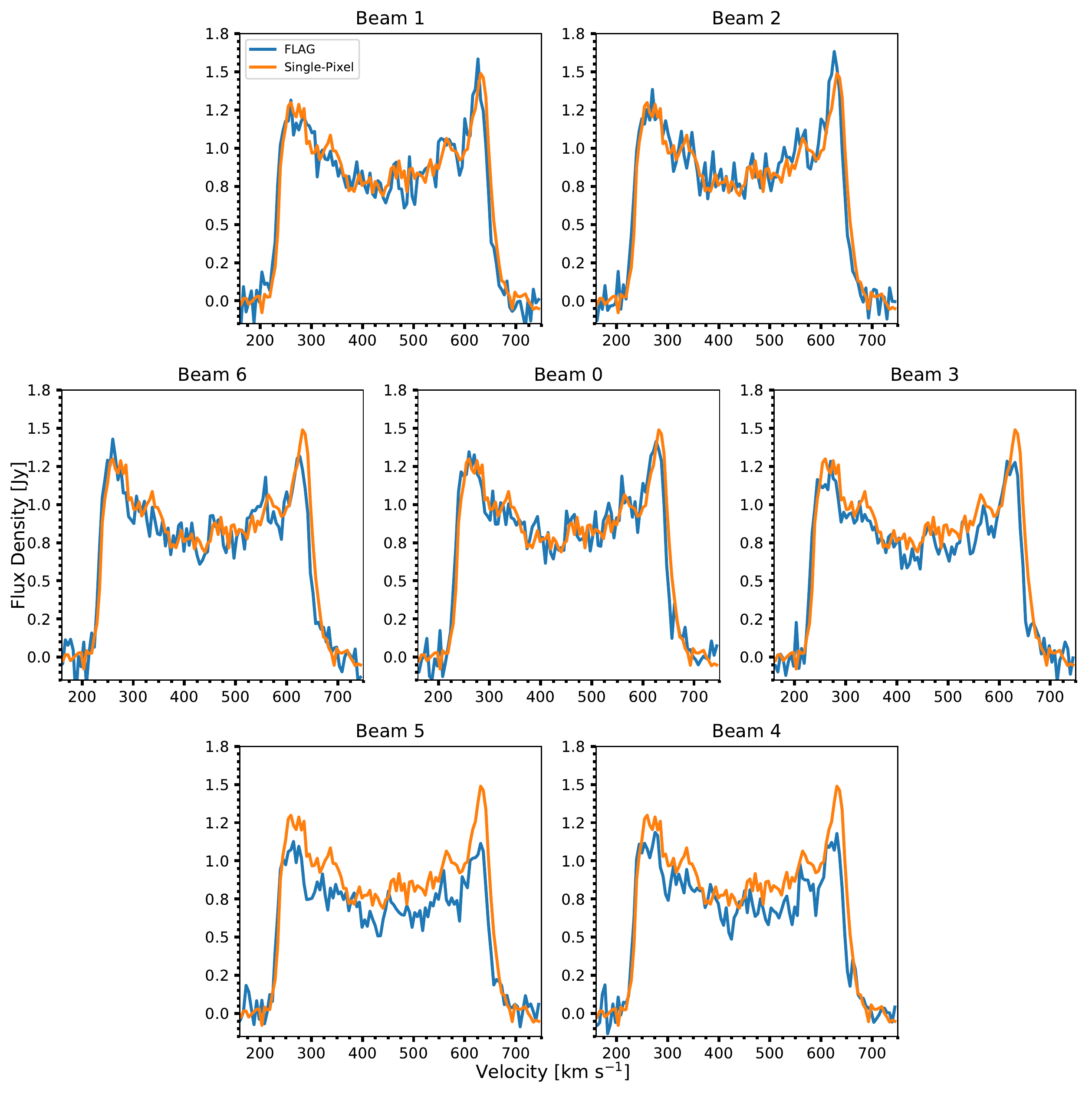}
\centering\caption{\label{fig:N4258FluxDen_IndvBeams} $\hi$ flux density profiles of NGC 4258 from each FLAG beam (blue) with equivalent single-pixel profile (orange) overlaid. These profiles were measured by integrating over the dashed rectangular region overlaided in  Figure~\ref{fig:N4258MomentMap}.}
\end{figure*}

\begin{figure*}
\includegraphics[width = \columnwidth]{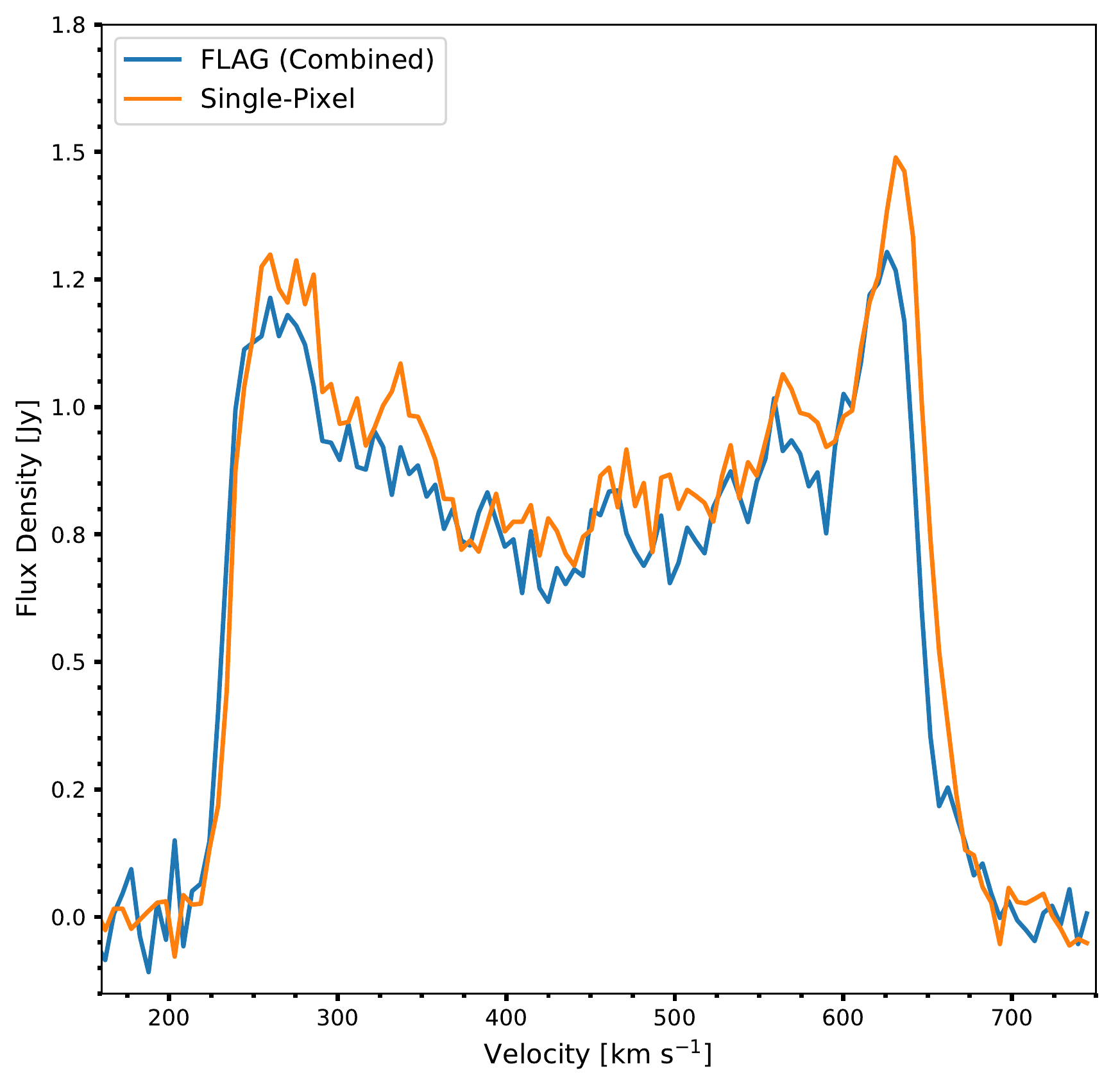}
\includegraphics[width = \columnwidth]{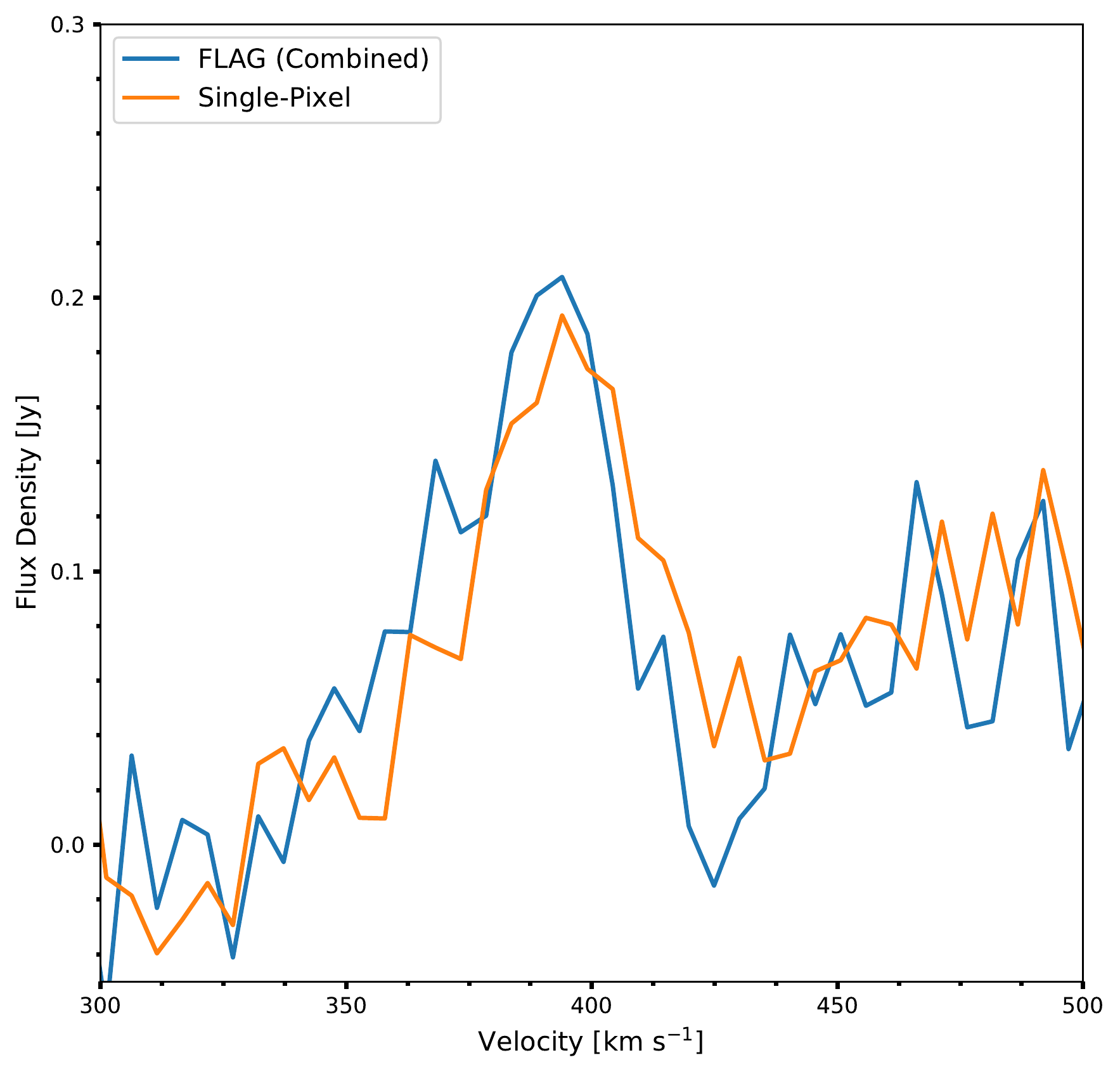}
\centering\caption{\label{fig:N4258FluxDen_CombBeams} Left: $\hi$ flux density profiles of NGC 4258 from the combined FLAG map (blue) with equivalent single-pixel profile (orange) overlaid. Right: $\hi$ flux density profiles from the same maps of the faint $\hi$ cloud, HIJASS J1219+46. These profiles were measured by integrating over the dashed and dot-dashed rectangular regions overlaid in  Figure~\ref{fig:N4258MomentMap}.}
\end{figure*}

Figures~\ref{fig:N4258FluxDen_IndvBeams} and \ref{fig:N4258FluxDen_CombBeams} present $\hi$ flux density profiles comparing FLAG and single-pixel data, with the former comparing measurements from individual beams and the latter showing profiles taken from the rectangular regions in the combined map as denoted in Figure~\ref{fig:N4258MomentMap}; the measured fluxes and associated $\hi$ masses are summarized in the row denoted $S_{\rm meas}$ under 17B\_360\_04 in Table~\ref{tab:summaryOfNoiseAndSS}. Since the intensity units of $\hi$ maps are presented in terms of surface brightness, it is vital to have knowledge of the beam area. Unfortunately, as demonstrated in the beam patterns shown in Figure~\ref{fig:beamMap}, each formed beam has a unique area. For each beam, we take the derived beam pattern and fit two Gaussians along two orthogonal cuts along the central horizontal and vertical axes. The beam area is then calculated from the average of these two Gaussians. The final beam area of the combined map is taken to be the mean of these individual beam areas; see again Table~\ref{tab:summaryOfNoiseAndSS} for a summary of these areas. Given the variation in early SEFD values, relative uncertainty with the final beam areas, possible errors in bandpass calibration, the presence of interference, and modeling for atmospheric effects, we adopt an overall 10\% flux uncertainty. 

The profiles and total flux measurements of Beams 0-3 and Beam 6 agree very well with the flux values from the equivalent single-pixel map. The offset in Beam 4 and Beam 5, while still within the 10\% flux uncertainty, is likely influenced by the deviations from Gaussianity in the main lobe of these formed beams and relatively high sidelobes. The combined maps and profiles of both NGC4258 and HIJASS J1219+46 agree very well with their single-pixel counterparts. The overall consistency between the FLAG and single-pixel data of the NGC 4258 Field and detection of a very diffuse $\hi$ cloud demonstrate the capability of FLAG to provide equivalent and accurate spectral line maps relative to the current single-pixel receiver on the GBT.  

%% FIGURE SHOWING FLUX PROFILES PER BEAM OF NGC4258

%% FIGURE SHOWING FLUX PROFILES OF COMBINED MAP FOR NGC4258 AND FILAMENT

\subsubsection{Galactic Center}\label{subsubsec:galacticCenter}
A recent single-pixel survey of $\hi$ above and below the Galactic Center undertaken by \citet{diTeodoro18} revealed a population of anomalous velocity clouds expanding out in a biconic shape, which likely arises from nuclear wind driven by the star formation activity in the inner regions of the Milky Way. As a demonstration of FLAG's capability to map extended Galactic emission and characterize gas moving at anomalous velocities, we mapped a 2$^{\circ}\times$2$^{\circ}$ region centered on $l$ = 353$^{\circ}$ and $b$ = $-$4$^{\circ}$ in the Galactic coordinate system with $\nu_{0}$ set to 1449.84841 MHz.

\begin{figure*}
\includegraphics[width = 3.5 in]{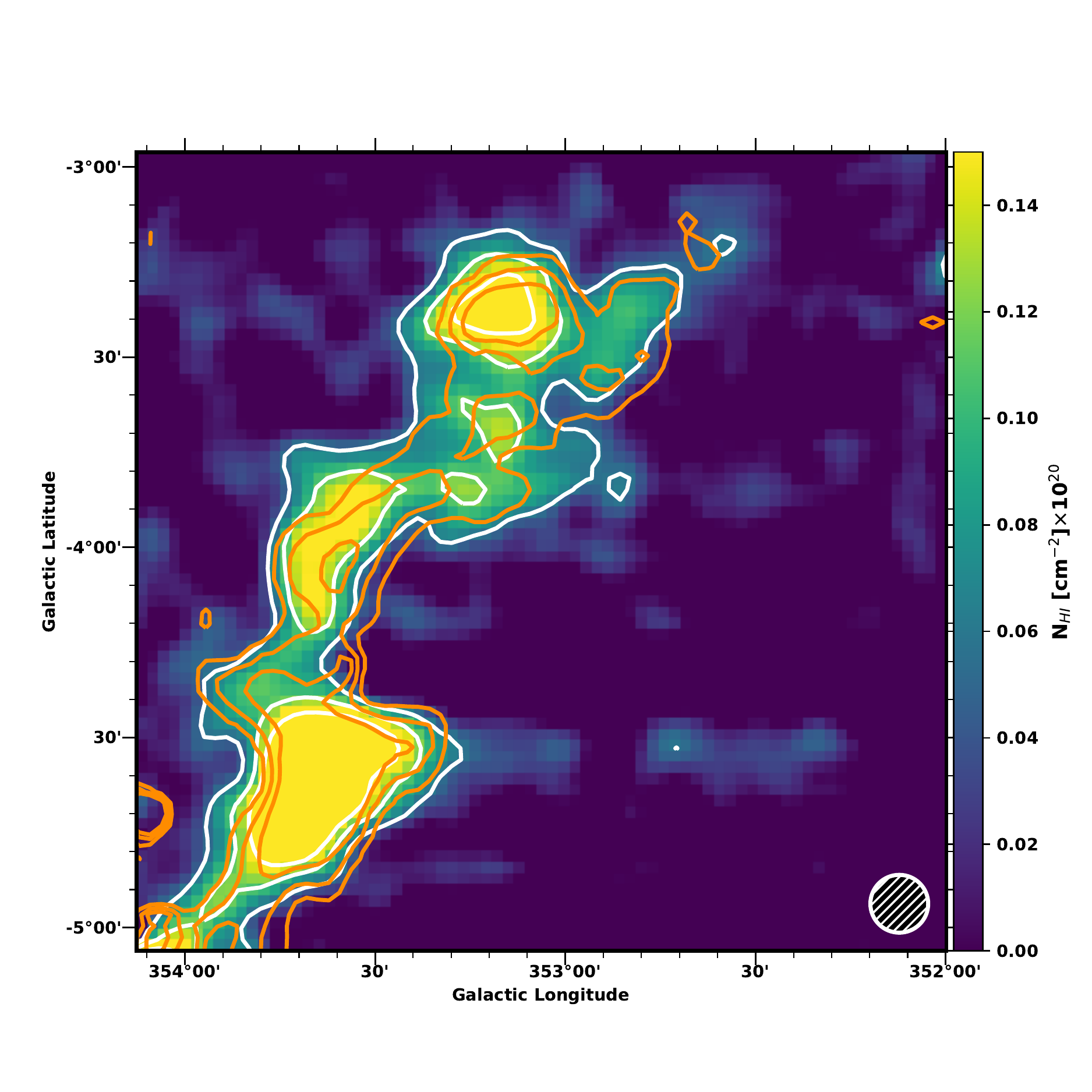}
\includegraphics[width = 3.5 in]{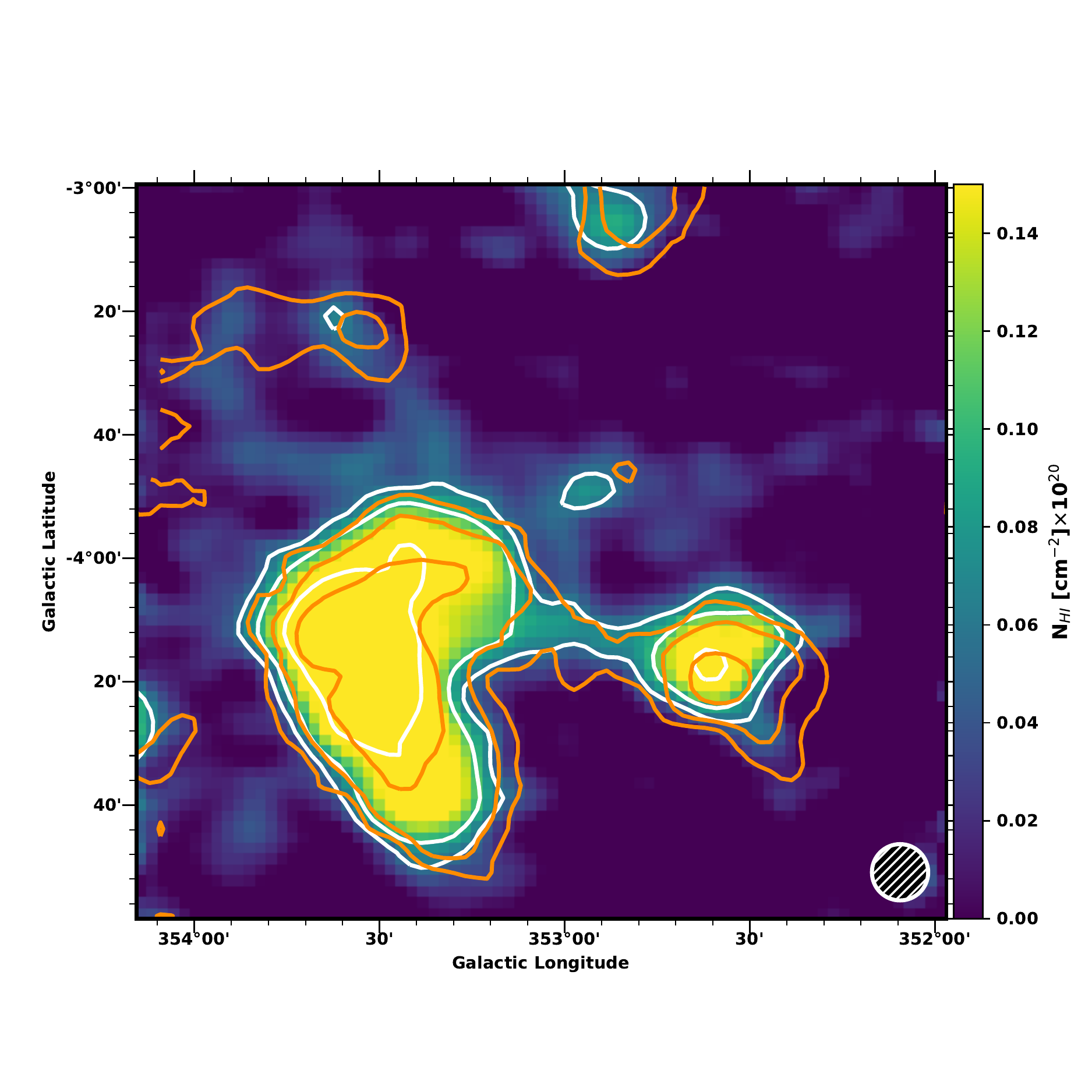}
\caption{\label{fig:GCMomentMaps} $\hi$ column density maps of the FLAG (color scale and white contours) with equivalent single-pixel data (orange contours) overlaid for the Galactic Center observations; \textit{left}: $\hi$ map derived by integrating over approaching LSR velocities (see text). The contours begin at a level of 6$\times$10$^{18}$ cm$^{-2}$ and continue at 5 and 10 times that level; \textit{right}: $\hi$ map derived by integrating over receding LSR velocities with the same contour levels.}  
\end{figure*}

Figure~\ref{fig:GCMomentMaps} presents $\hi$ column density of structures towards the Milky Way center that are moving at anomalous approaching and receding velocities. Once more, the spatial distribution of the emission detected by FLAG is sufficiently consistent with the single-pixel contours. The comparisons of $\hi$ spatial extent clearly highlight FLAG's ability to characterize both the diffuse $\hi$ associated with extragalactic sources and the complex kinematic properties of anomalous velocity clouds in and around the Milky Way.

%% FIGURE SHOWING MOM0 MAPS OF GC

%% SUBSECTION ON DISCREPANCIES BETWEEN FLAG AND SP MAPS
\subsubsection{Discrepancies and Improvements}\label{subsubsec:discrepancies}

%% FIGURE SHOWING BEAM0/BEAM2 BEAM PATTERN WITH CONTOURS FROM SP MODEL
\begin{figure}
\includegraphics[width = \columnwidth]{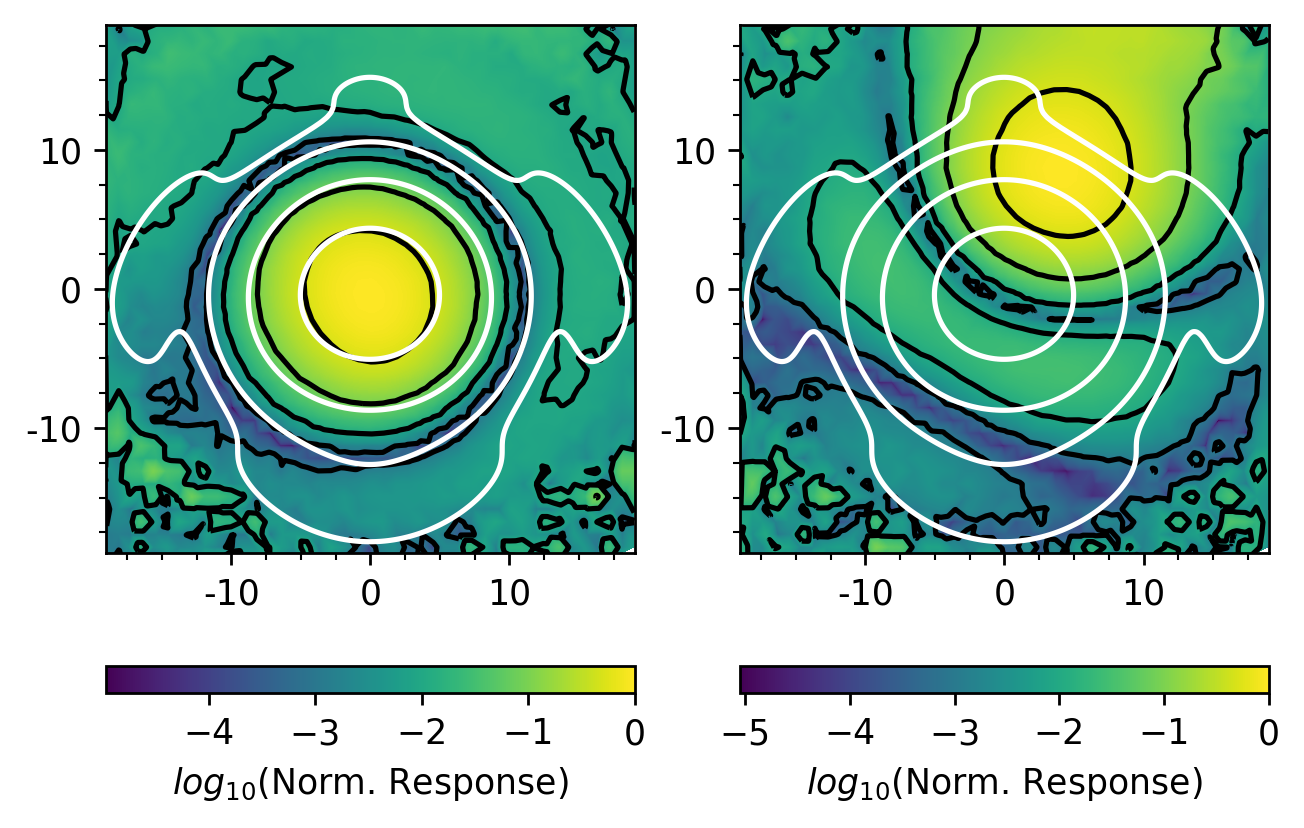}
\caption{\label{fig:beam0_2_SP_Contour}The beam patterns from the boresight FLAG beam (left) and outer Beam 2 (right). The white contours denote the response from a model of the singel-pixel beam. Contours begin at a level of 0.001, 0.01, 0.1 and 0.5 times the peak response. The outer beam is shown to highlight the highly peaked sidelobe that overlaps near the peak of the boresight.}
\end{figure}

Figures~\ref{fig:N6946ChanMaps}-\ref{fig:GCMomentMaps} demonstrate broad agreement with previous single-pixel observations. However, there are notable discrepancies between FLAG and single-pixel contours that are at the same absolute flux density and column density levels. There are several possible sources for such discrepancies including stray radiation from the complex beam shapes, differences in sensitivity between maps, and a flux offset between FLAG and single-pixel data.

Figure~\ref{fig:beam0_2_SP_Contour} shows the beam patterns for the boresight (Beam 0) and Beam 2 derived from the calibration grid from session GBT17B\_360\_04 with overlaid contours from a model of the GBT single-pixel L-Band beam shown in Figure 1 of ~\citet{pingel18}. There is excellent agreement between the single-pixel beam model and Beam 0 from the FWHM response level extending down to the level of the first sidelobe at the 0.1\%. The sidelobes is highly asymmetric in both FLAG beams, with the peak sidelobe in the outer Beam 2 peaking an order of magnitude higher than that of the boresight beam; also, note that this sidelobe overlaps almost directly with the peak of the boresight response. Given that that dynamic range of the our observations is typically on the order of several hundred, it is feasible that such complex beam shapes --- especially in the final combined FLAG maps, where the beam responses are effectively averaged together -- will affect the observed morphology of diffuse structures.

%% FIGURE SHOWING MOM0 MAP OF CONVOLVED NGC6946 DATA
\begin{figure}
\includegraphics[width = \columnwidth]{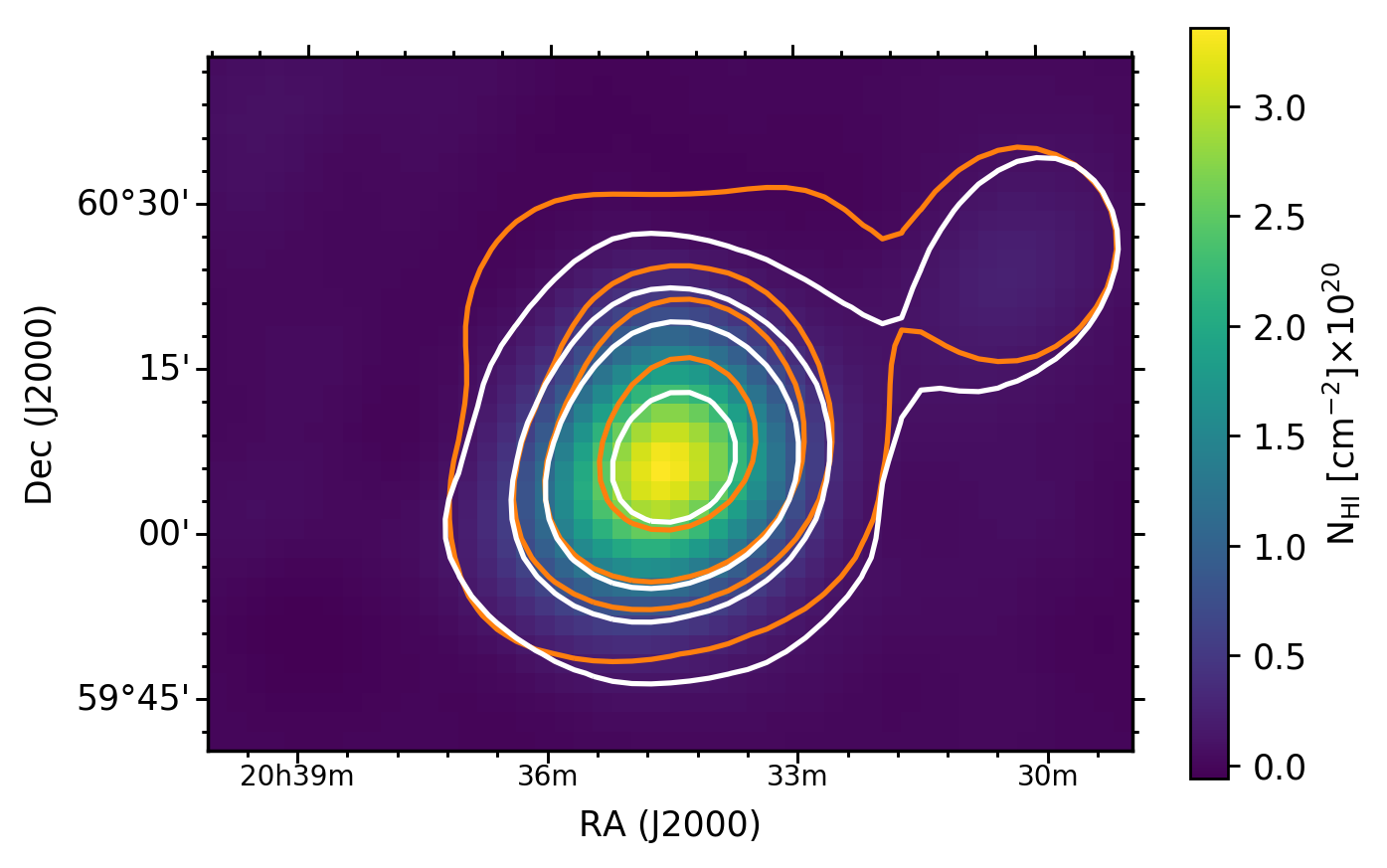}
\caption{\label{fig:N6946Mom0Map_Conv} $\hi$ column density map of NGC 6946 after convolving the FLAG data with a model of the GBT L-Band single-pixel beam (color scale and white contours) with contours from the data single-pixel overlaid after a similar convolution with the FLAG boresight beam pattern. The convolution ensures the both maps have effectively equal responses to the observed sky brightness distribution. The contours are at the same levels listed in the caption from Figure~\ref{fig:N6946Mom0Map}.}
\end{figure}

To test the degree to which the complex sidelobes structure in the formed FLAG beams affect the discrepancies in the flux density contours, we convolve the FLAG map made with the boresight beam of NGC 6946 with a single-pixel beam model re-gridded to a common pixel grid. Likewise, the single-pixel map is convolved with the FLAG boresight beam pattern; the resulting column density map shown in Figure~\ref{fig:N6946Mom0Map_Conv} now shows the same sky brightness distribution convolved with the same response. The apparent bridge of material that now connects NGC 6946 with its companions is due to the degraded angular resolution from convolution with both beams. The contours to the south are in better agreement with deviations on the scale of a single pixel, confirming that the asymmetric sidelobe patterns of the formed FLAG beams indeed influence the morphology of diffuse emission by a non-negligible amount. The larger discrepancies towards the north and around the unresolved companions can be attributed to the order of magnitude difference in sensitivity between the FLAG and single-pixel map, which detects an appreciable amount of diffuse $\hi$ below a column density level of 1$\times$10$^{19}$ cm$^{-2}$ --- including a conspicuous $\hi$ plume --- that likely influences these northern contours \citep{pisano14}. Differences between the overall flux scale, which has since been addressed with improvements to the overall stability of the system, can also cause such discrepancies.

There are several possible avenues to mitigate effects from the complex sidelobe patterns, including utilizing alternative beamforming algorithms. However, attempts to constrain the sidelobe levels of formed beams on other radio telescopes sacrifice sensitivity at unacceptable levels. Fortunately, the raw covariance data obtained from FLAG can be used to aid development of new algorithms. A more traditional approach would be to apply a stray radiation correction first developed by \citet{vanWoerden1962}, demonstrated for single dish telescopes e.g., \citet{kalbera1980a}, \citet{winkel2016}, and applied to multibeam systems in \citet{kalberla2010}. Such a correction requires detailed knowledge of the sidelobes, which can easily be obtained using a sufficiently large calibration grid. The correction can also be considerably simplified by having a known all-sky brightness temperature distribution. Ample archival data from the single-pixel exists to attempt such corrections for future FLAG data.

%% SUBSECTION ON SURVEY SPEED METRIC & COMPARISON
\subsection{Survey Speed Comparison}\label{subsec:surveySpeed}

We now aim to quantify the performance of FLAG relative to the single-pixel receiver and the PAFs and multi-beam receivers available on other prominent radio telescopes. We do this through the survey speed (SS) metric.

To obtain an expression for $SS$, we first define a given surface brightness sensitivity (in units of K) to be
%% EQUATION FOR SURFACE BRIGHTNESS SENSITIVITY
\begin{equation}\label{eq:surfbrightSens}
\sigma = \frac{T_{\rm sys}}{\sqrt{\Delta\nu N_{\rm p} t}}, 
\end{equation}
where $\Delta\nu$ is the width of a frequency bin, $N_{\rm p}$ is the number of polarizations, and $t$ is the integration time necessary to reach a given surface brightness sensitivity. Putting $T_{\rm sys}$ in terms of SEFD (Equation~\ref{eq:Ta_eta}) and absorbing the antenna gain factors gives an equivalent expression for point source sensitivity ($\sigma_{\rm s}$ in units of Jy) that can be rearranged to give the time necessary to reach a given point source sensitivity
%% EQUATION FOR TIME NECESSARY TO REACH POINT-SOURCE SENSITIVITY
\begin{equation}\label{eq:effectiveIntTime}
t = \frac{1}{\Delta\nu N_{\rm p}}\left(\frac{\sigma_{\rm s}}{\rm SEFD}\right)^2
\end{equation}

Following \citet{johnston2006}, the speed at which a single dish can survey an area of sky to the necessary sensitivity limit is the ratio of its inherent FoV to $t$ or 
%% EQUATION FOR SS
\begin{equation}\label{eq:ss}
SS = {\rm FoV}\Delta\nu N_{\rm p}\left(\frac{\sigma_{\rm s}}{\rm SEFD}\right)^2
\end{equation}
where the FoV is measured in square degrees. In the case of FLAG, we define the FoV to be the area of sky over which the sensitivity map (see again Figure~\ref{fig:sensMap}) remains above a $-$3 dB drop off relative to peak response. The average FoV measured from all available calibration grids is 0.144 deg$^2$.

We employ the Source Finding Application (SoFiA; \citealt{serra2015_SoFiA}) software package to measure the noise in the FLAG cubes and compare with similar data from the single-pixel receiver. We utilize the feature in which the rms is estimated from a Gaussian fit to the negative half of the histogram of pixel values. The histogram is constructed using only emission-free channels to avoid spectral channels whose reference spectra have been contaminated by Milky Way emission during calibration. Table~\ref{tab:summaryOfNoiseAndSS} lists the measured noise returned by SoFiA for the cubes produced for each individually formed beam, the combined beam cube, and the single-pixel cube. The measured noise in the combined beam cubes generally scale by the reciprocal of the square root number of beams, as expected from pure Gaussian noise. The beam-to-beam variation in SEFD values also influences the final noise floor in the combined cubes. Because the available single-pixel cubes are generally more sensitive than the FLAG commissioning maps, a straight calculation of the $SS$ metric using the measured noise properties will give a convoluted comparison. To ensure a normalized comparison, we use the measured single-pixel noise while taking the FLAG SEFD values available in Table~\ref{tab:SEFDSummary} to compute Equation~\ref{eq:ss}. The quoted uncertainties are propagated from the SEFD uncertainties. For all observing sessions, FLAG possesses a higher $SS$ in the final combined maps, largely aided by the increase in FoV.

\begin{figure}
\includegraphics[width=\columnwidth]{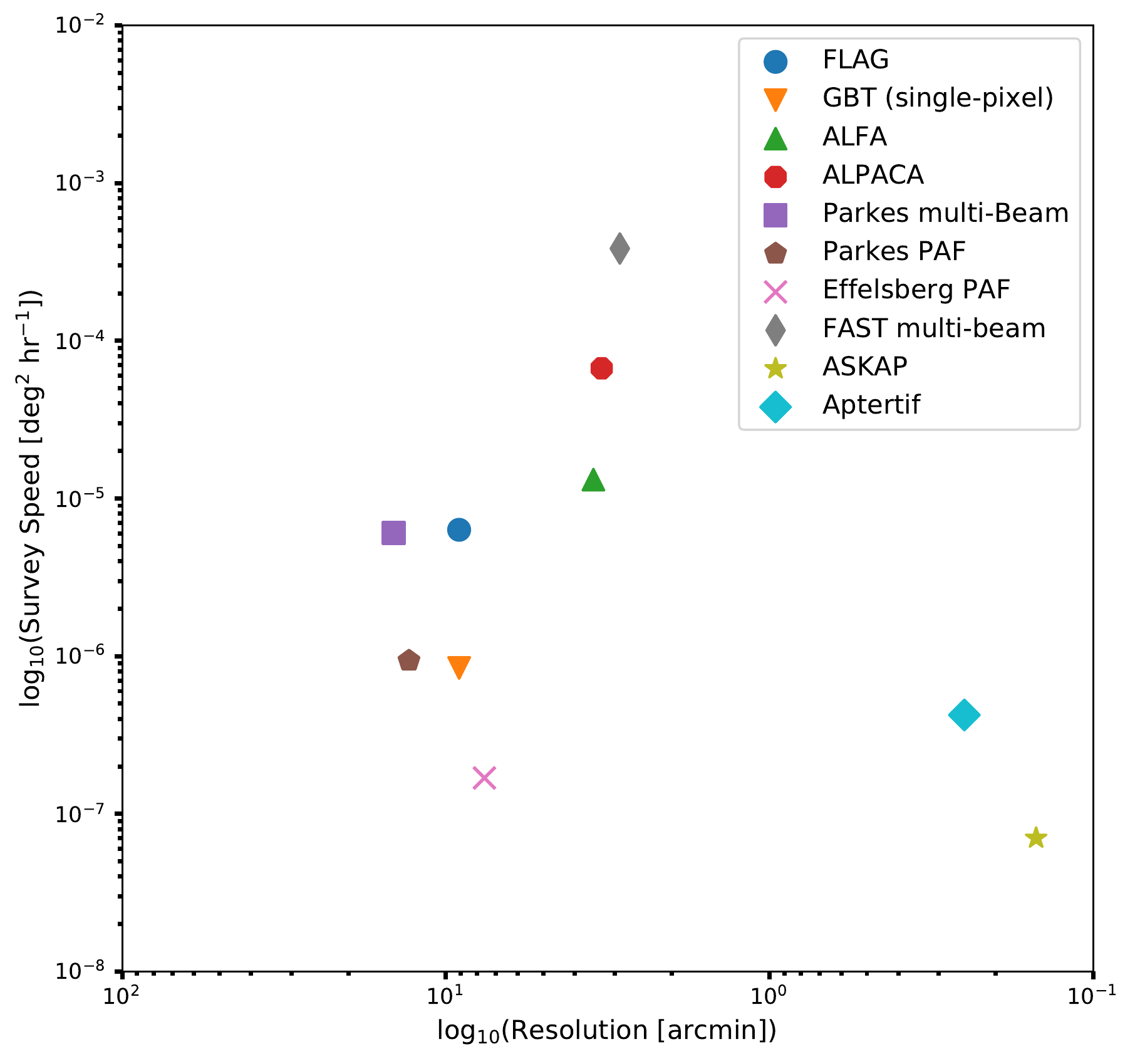}
\caption{\label{fig:surveySpeedComparison} Comparison of various receiver survey speeds. The dotted lines denote different PAF recievers, while the solid lines represent traditional multi-beam and single-pixel receivers.}
\end{figure}

As broader comparison, we assume a desired point source sensitivity level of 5 mJy and plot the SS of FLAG, the single-pixel receiver, and several other multi-pixel receivers and PAFs already available or planned for other major radio telescopes as function of angular resolution in Figure~\ref{fig:surveySpeedComparison}. When comparing different receivers, we must make a consistent definition of the FoV, since sensitivity maps for the other receivers are not readily available. In these cases, we consider the field of view to be
%%EQUATION FOR FOV_eff
\begin{equation}\label{eq:fovEff}
%{\rm FoV}_{\rm eff} = N_{\rm b}\beta\Omega_{\rm b},
{\rm FoV}_{\rm eff} = N_{\rm b}\Omega_{\rm b}, 
\end{equation}
where $N_{\rm b}$ is the number of beams and $\Omega_{\rm b}$ is the beam solid angle in square degrees as measured at the FWHM.
%, and $\beta$ represents some beam overlap factor to account for cases when the beams are spaced wider than Nyquist spacing ($d_{\rm Nyq}$ = $\frac{\lambda}{2D}$. If the beams are separated by some angular distance $d$, we define
%% EQUATION FOR BETA
%\begin{equation}\label{eq:beta}
%\beta = \frac{d}{d_{\rm Ny}}. 
%\end{equation}
%We can now compare $SS$ metrics of different receivers when an unambiguous definition of the FoV is not available using a sensitivity map while also taking into account the scenario when multiple passes on a similar region of sky to fully sample. In the case of FLAG, the beams are intentionally placed very near the Nyquist spacing, thus $\beta$$\sim$1. 
Table~\ref{tab:surveySpeedParams} summarizes the parameters used in the calculation of Equation~\ref{eq:ss}. 

ASKAP and Apertif, being PAF-equipped interferometric telescopes, possess a distinct advantage in terms of angular resolution due to their capability to sample large spatial frequencies. However, even when considering point-source sensitivity, they are ultimately limited in their SS by their relatively large SEFDs. On the other hand, the SEFDs of single dish telescopes benefit from their large and continuous apertures but suffer in terms of angular resolution. The SS of FLAG relative to the GBT single-pixel reciever is about an order of magnitude higher, and the cryogenically cooled LNAs in its front end enhance its performance to exceed all other existing PAFs, while providing comparable resolution. Relative to multiple horn receivers, FLAG beats the 13 beam multibeam receiver on Parkes in terms of angular resolution and SS and also produces comparable SS metrics to the 7-beam ALFA receiver on the now defunct 300m Arecibo telescope. In fact, the survey capabilities of the GBT when equipped with FLAG are only exceeded by the multibeam receiver on FAST, the world's largest primary reflector telescope that cannot be fully steered.

\begin{table*}
\centering
\resizebox{\textwidth}{!} 
{\begin{tabular}{llcccccccccc}
\hline \hline
\\[-1.0em]

Session           &   Property    & Beam 0     & Beam 1     & Beam 2     & Beam 3     & Beam 4     & Beam 5    & Beam 6    & Combined & single-pixel \\ \hline 
\\[-1.0em]
\textbf{16B\_400\_12}             &            &            &            &            &            &            &            &             &            \\
% & $S_{\rm meas}$ [Jy km s$^{-1}$]& 430$\pm$90 & 340$\pm$70 & 410$\pm$80 & 460$\pm$90 & 460$\pm$90 & 410$\pm$80 & 410$\pm$80 & 410$\pm$80  & 440$\pm$20 \\
 & $\sigma_{\rm meas}$ [mJy Beam$^{-1}$]       & 43         & 45         & 46         & 46         & 49            & 49         & 49          & 19 & 4     \\
 & SS [deg$^2$ hr$^{-1}$]            & 0.38$\pm$0.05  & 0.3$\pm$0.2   & 0.3$\pm$0.2   & 0.32$\pm$0.04   & 0.29$\pm$0.04      & 0.3$\pm$0.1   & 0.3$\pm$0.1  & 1.70$\pm$0.08 & 0.78$\pm$0.05 \\
\textbf{16B\_400\_13}             &            &          &          &          &          &           &           &             &               \\
% & $S_{\rm meas}$ [Jy km s$^{-1}$]& 370$\pm$70 & 300$\pm$60 & 340$\pm$70 & 340$\pm$70 & 250$\pm$50 & 300$\pm$60 & 300$\pm$60 & 310$\pm$60 & 440$\pm$20 \\
 & $\sigma_{\rm meas}$ [mJy/Beam$^{-1}$] & 44        & 49       & 46       & 51       & 53       & 57        & 51        & 20          & 4             \\
 & SS [deg$^2$ hr$^{-1}$]          & 0.37$\pm$0.05 & 0.3$\pm$0.2 & 0.3$\pm$0.2 & 0.32$\pm$0.04 & 0.29$\pm$0.04 & 0.3$\pm$0.1 & 0.3$\pm$0.1 & 1.70$\pm$0.08    & 0.78$\pm$0.05 \\
\textbf{17B\_360\_03}             &           &          &          &          &          &           &           &             &               \\
% & $S_{\rm meas}$ [Jy km s$^{-1}$]& 700$\pm$100 & 700$\pm$100 & 700$\pm$100 & 700$\pm$100 & 700$\pm$100 & 600$\pm$100 & 700$\pm$100 & 700$\pm$100 & 410$\pm$20 \\
& Scaling Factor{\dag} & 0.58 & 0.57 & 0.60 & 0.53 & 0.50 & 0.57 & 0.56 & & \\ 
 & $\sigma_{\rm meas}$ [mJy/Beam$^{-1}$] & 30        & 31       & 29       & 30       & 33       & 33        & 33        & 16          & 8             \\
 & SS [deg$^2$ hr$^{-1}$]          & 1.09$\pm$0.01  & 1.01$\pm$0.02 & 1.11$\pm$0.05 & 1.02$\pm$0.04 & 0.89$\pm$0.02 & 0.97$\pm$0.02  & 0.96$\pm$0.02  & 5.56$\pm$0.02 & 3.1$\pm$0.2   \\
 \textbf{17B\_360\_04}            &           &          &          &          &          &           &           &             &               \\
 & $\Omega$ [arcmin$^2$]          & 95 & 100 & 101 & 105 & 110 & 122 & 114 & 107 & 94 \\  
 & $S_{\rm meas}$ [Jy km s$^{-1}$]& 410$\pm$40 & 400$\pm$40 & 420$\pm$40 & 370$\pm$40 & 350$\pm$40 & 340$\pm$30 & 390$\pm$40 & 380$\pm$40 & 410$\pm$20 \\
 & $\sigma_{\rm meas}$ [mJy/Beam$^{-1}$] & 15        & 15       & 15       & 15       & 16       & 15        & 15        & 8           & 8             \\
 & SS [deg$^2$ hr$^{-1}$]          & 3.38$\pm$0.02  & 3.24$\pm$0.06 & 3.17$\pm$0.03 & 3.17$\pm$0.06 & 3.11$\pm$0.06 & 3.24$\pm$0.06  & 3.11$\pm$0.03  & 17.74$\pm$0.04 & 3.1$\pm$0.2   \\

\\ [-1.0em] \hline
\end{tabular}}
\caption{Measured noise ($\sigma_{\rm meas}$), survey speeds ($SS$), beam area ($\Omega$), and measured flux ($S_{\rm meas}$); {\dag} represents the scaling factor applied before combination with an associated frequency-dithered session.}
\label{tab:summaryOfNoiseAndSS}
\end{table*}

%% TABLE SUMMARIZING SURVEY SPEED CALCULATION FOR DIFFERENT RECEIVERS
\begin{table*}
\centering
\resizebox{\textwidth}{!}
{\begin{tabular}{lccccccc}
\hline \hline
\\[-1.0em]
Receiver & $N_{\rm b}$ & FWHM [arcmin] & Resolution [arcmin] & FoV$_{\rm eff}$ [deg$^2$] & SEFD [Jy] & SS [deg hr$^{-1}$]  & Reference \\ \hline
\\[-1.0em]
FLAG                         & 7    & 9.1  & 9.1  & 0.144 & 10 & 6.3$\times$10$^{-6}$ & This work  \\
GBT single-pixel             & 1    & 9.1  & 9.1  & 0.018 & 9.7 & 8.4$\times$10$^{-7}$ & This work \\
Apertif                    & 37   & 30.0 & 0.3 &  10.500  & 330 &  4.2$\times$10$^{-7}$ & \citealt{oosterloo09} \\ 
ASKAP                        & 36   & 60.0 & 0.2 &  46.200  & 1700 & 7.0$\times$10$^{-8}$ & David McConnell (2020; private communication) \\
ALFA                         & 7    & 3.5  & 3.5  & 0.027 & 3 & 1.3$\times$10$^{-5}$ & \citealt{peek2011}; \url{http://outreach.naci.edu/ao/scientist-user-portal/astronomy/recievers} \\
ALPACA                       & 40   & 3.3  & 3.3  & 0.137 & 3 &  6.7$\times$10$^{-5}$ & \citealt{roshi2019}  \\
Effelsberg PAF               & 36   & 7.6  & 7.6  & 0.650 & 130 & 1.7$\times$10$^{-7}$ & \citealt{rajwade2019} \\
FAST Mutli-Beam              & 19   & 2.9  & 2.9  & 0.014 & 0.4  & 3.8$\times$10$^{-4}$ & \citealt{Li2020} \\
Parkes Multi-Beam            & 13   & 14.5 & 14.5 & 0.86 & 25 & 6.0$\times$10$^{-6}$ & \citealt{staveley-smith1996, mcclure-griffiths2009}  \\
Parkes PAF                   & 17   & 13.0 & 13.0 & 0.900 & 65 & 9.4$\times$10$^{-7}$ & \citealt{Reynolds17} \\

\\ [-1.0em] \hline
\end{tabular}}
\caption{Survey Speed Parameters. Note that the FWHM for ASKAP and Apertif refer to the size of a single formed primary beam, while resolution refers to the size of a typical synthesized beam.}
\label{tab:surveySpeedParams}
\end{table*}

%% FIGURE SHOWING SURVEY SPEED COMPARISON OF DIFFERENT RECEIVERS

%% CONCLUSIONS AND FUTURE OF INSTRUMENT
\section{Conclusions and Outlook}\label{sec:conc}
This work summarized the commissioning of the calibration and spectral-line observing modes for a new beamforming back end for FLAG, a cryogenically cooled PAF L-band receiver for the GBT. These observations represent the culmination of several commissioning runs from 2016 to 2018 wherein the system was incrementally tested on a diverse range of extragalactic and Galactic $\hi$ science targets and known calibrator sources. The main results from these commissioning runs are:
\begin{itemize}
    \item The beamforming weights derived from Calibration Grids and 7Pt-Cal scans produce seven simultaneously formed beams optimally spaced to achieve uniform sensitivity across the FoV. The measured beam shapes are sufficiently Gaussian down to the 3\% level of the peak response with FWHM's ranging from 8.7$'$ to 9.5$'$. The locations of the peak response for each beam beam are reliably located within 5\% of the their intended pointing centers.  
    \item The custom python package, {\tt pyFLAG}, is used to apply the beamforming weights to the raw covariance matrices to create SDFITS files that contain uncalibrated beamformed spectra. Through several GBTIDL and GBO tools, these spectra are flux calibrated and imaged to create SDFITS cubes for each formed beam. A beam combined cube is produced by averaging all spectra from these individual cubes. 
    \item  The overall phase of the derived complex beamforming weights varies less than 1\% over timescales of $\sim$1 week, indicating the directional response to identically coincident signals is extremely reliable. Applying stale weights (i.e., weights from a previous observing session) to the steering vectors of a subsequent observing session produces beams that keep their Gaussian shape above the 50\% level of the peak response, but degrades the side-lobe structure, sensitivity, and shifts the peak response away from the intended pointing center. An observer should at least perform a 7Pt-Cal scan to derive contemporaneous weights. In the future, the word lock calibration procedure will ensure the phase response of a previous set of weights applies to the current state of the system. Weights can then be reused without deterioration of sensitivity or overall beam shape.
    \item The measures of sensitivity across the entire 150 MHz bandpass show steady improvement over our commissioning runs. Likewise, the measured SEFDs used to scale spectra to the correct flux scale converged towards the single-pixel value in later sessions. These improvements are the result of improvements in our calibration strategies to obtain and maintain bit and byte-lock, which ensure the serialized complex voltages samples streaming from the front end over optical fiber are correctly decoded for downstream processing in the back end.    
    \item The observed $\hi$ science targets were chosen to incrementally test the spectral line mapping capabilities of FLAG. The map of NGC 6946 compares well with equivalent single-pixel data. The $\hi$ flux density profiles of sources within the NGC 4258 field are also well-matched to equivalent single-pixel data and demonstrate accurate measurements of the shape of the FLAG beams. The detection of the diffuse $\hi$ cloud,  HIJASS J1219+46, and emission at anomalous velocities towards the Galactic Center shows that FLAG is able to reproduce a wide-range of $\hi$ properties observed in and around extragalactic sources and Galactic regions.
    \item The relatively high sidelobes inherent to maxSNR beamforming do affect the overall morphology of low-level emission. Correcting for stray radiation using proven techniques can mitigate these effects in future observations.
    \item The compromise between survey speed and angular resolution when compared between FLAG, the current GBT single-pixel receiver, and other multi-beam and PAF receivers available or planned for the world's major radio telescopes is only matched by those with much larger apertures that are not fully steerable. 
\end{itemize}

Overall, the new beamforming back end for FLAG performed exceptionally well in terms of the derivation of stable beamforming weights and generally reproduces equivalent observations from the current single-pixel receiver. There are several possible avenues of improvement including the correcting for stray radiation. The increase in survey speed provided by FLAG and its upgraded backend, coupled with the sky coverage available only from a fully steerable dish, will ensure the GBT remains a premiere instrument for radio astrophysics.

\acknowledgments{The authors wish to thank Richard Prestage for leading the organizational efforts during these commissioning observations and for significant contributions to the field of radio astronomy. We also thank the anonymous referee whose comments greatly improved the quality of this work. We acknowledge the significant funding for the FLAG receiver provided by GBO and NRAO. The Green Bank Observatory is a major facility supported by the National Science Foundation and operated under cooperative agreement by Associated Universities, Inc. The National Radio Astronomy Observatory is a facility of the National Science Foundation operated under cooperative agreement by Associated Universities, Inc. NMP, KMR, DRL, DA, DJP, and MAM acknowledge partial support from National Science Foundation grant AST-1309815. KMR acknowledges funding from the European Research Council (ERC) under the European Union's Horizon 2020 research and innovation programme (grant agreement No 694745). This material is based upon the work supported by National Science Foundation Grant No. 1309832.}
%\vspace{1mm}


\software[htp]{This research made use of Astropy,\footnote{\url{http://www.astropy.org}} a community-developed core Python package for Astronomy \citep{astropy:2013, astropy:2018}.}



\vspace{65mm}
\bibliography{bibFile}{}
\bibliographystyle{aasjournal}



\end{document}